\documentclass[useAMS,usenatbib, usegraphicx]{mn2e}

\usepackage{epsfig}
\usepackage{bm}
\usepackage{amsmath}
\usepackage{amssymb}
\usepackage{natbib}
\newcommand{\kms}{\ensuremath{{\rm km\,s}^{-1}}}
\newcommand{\msun}{\ensuremath{{\rm M}_{\odot}}}
\newcommand{\rsun}{\ensuremath{{\rm R}_{\odot}}}

\newcommand{\yr}{\ensuremath{\rm yr}}
\newcommand{\myr}{\ensuremath{\rm Myr}}
\newcommand{\msmbh}{\ensuremath{M_{\rm SMBH}}}
\newcommand{\infinity}{{\infty}}

\newcommand{\sag}{Sgr~A*}
\newcommand{\rmd}{{\rm d}}

\newcommand{\mtot}{\ensuremath{M_{\rm tot}}}

\newcommand\lsim{\mathrel{\rlap{\lower4pt\hbox{\hskip1pt$\sim$}}
        \raise1pt\hbox{$<$}}}
\newcommand\gsim{\mathrel{\rlap{\lower4pt\hbox{\hskip1pt$\sim$}}
        \raise1pt\hbox{$>$}}}
\newcommand\propsim{\mathrel{\rlap{\lower4pt\hbox{\hskip1pt$\sim$}}
        \raise1pt\hbox{$\propto$}}}

\newcommand{\D}{\mathrm{d}}

\newcommand{\Hz}{\,\mathrm{Hz}}

\newcommand{\Msun}{\mathrm{M}_{\odot}}

\newcommand{\baker}{2007PhRvD..75l4024B}
\newcommand{\bc}{2004PhRvD..69h2005B}
\newcommand{\berti}{2007PhRvD..76j4044B}

\newcommand{\gairof}{2005PhRvD..72h4009G}
\newcommand{\gairos}{2006PhRvD..74j9901G}
\newcommand{\hindera}{2008PhRvD..77h1502H}
\newcommand{\hinderb}{2008arXiv0806.1037H}
\newcommand{\pmat}{1963PhRv..131..435P}
\newcommand{\pk}{2007CQGra..24...83P}
\newcommand{\peters}{1964PhRv..136.1224P}
\newcommand{\w}{2008arXiv0802.2520W}

\newcommand{\tu}{1977ApJ...216..610T}

\newcommand{\morris}{1993ApJ...408..496M}
\newcommand{\meg}{2000ApJ...545..847M}
\newcommand{\new}{}

\title[GWs from BHs in Galactic Nuclei]{Gravitational waves from scattering of stellar-mass black holes in galactic nuclei}

\author[O'Leary, Kocsis,\& Loeb]{Ryan M.\
O'Leary$^{1}$\thanks{E-mail:roleary@cfa.harvard.edu}, Bence Kocsis$^{1.2}$\thanks{E-mail:bkocsis@ias.edu}, and Abraham
Loeb$^{1}$\thanks{E-mail:aloeb@cfa.harvard.edu}\\$^{1}$Harvard-Smithsonian Center for Astrophysics, 60 Garden St., Cambridge, MA 02138, USA\\$^{2}$Institute for Advanced Study, Einstein Drive, BH-151, Princeton, NJ 08540}

\begin{document}
\maketitle

\begin{abstract}
  Stellar mass black holes (BHs) are expected to segregate and form a
  steep density cusp around supermassive black holes (SMBHs) in
  galactic nuclei.  We follow the evolution of a multi-mass system of
  BHs and stars by numerically integrating the Fokker-Planck energy
  diffusion equations for a variety of BH mass distributions. We find
  that the BHs ``self-segregate'', and that the rarest, most massive
  BHs dominate the scattering rate closest to the SMBH ($\lesssim
  10^{-1}\,$pc).  BH--BH binaries form out of gravitational wave
  emission during BH encounters. We find that the expected rate of BH
  coalescence events detectable by Advanced LIGO is {\new $\sim
    1-10^2\,$yr$^{-1}$, depending on the initial mass function of
    stars in galactic nuclei and the mass of the most massive BHs. We
    find that the actual merger rate is likely $\sim 10$ times larger
    than this due to the intrinsic scatter of stellar densities in
    many different galaxies.}  The BH binaries that form this way in
  galactic nuclei have significant eccentricities as they enter the
  LIGO band ($90\%$ with $e > 0.9$), and are therefore distinguishable
  from other binaries, which circularise before becoming
  detectable. We also show that eccentric mergers can be detected to
  larger distances and greater BH masses than circular mergers, up to
  $\sim 700\,\Msun$.  Future ground-based gravitational wave
  observatories will be able to constrain both the mass function of
  BHs and stars in galactic nuclei.
\end{abstract}

\begin{keywords}
galaxies:kinematics and dynamics--galaxies:nuclei--black hole
physics--gravitational waves
\end{keywords}

\section{Introduction}
\subsection{Motivation}\label{sec:motivation}
The coalescence of stellar mass black hole-black hole (BH-BH) binaries
is one of the most anticipated sources of gravitational waves for
ground-based interferometers such as LIGO\footnote{http://www.ligo.caltech.edu/} or VIRGO\footnote{http://www.virgo.infn.it/}.
Whether formed primordially \citep{2002ApJ...572..407B, 2004ApJ...611.1068B,
2008ApJ...676.1162S}
or in dense star clusters \citep{2000ApJ...528L..17P, 2004ApJ...616..221G,
2006ApJ...637..937O, 2006ApJ...640..156G, 2007PhRvD..76f1504O,
2008arXiv0804.2783M}, nearly all of these binaries are expected to be
circularised by gravitational wave (GW) emission before they are detectable
by ground based observatories \citep[][]{2006ApJ...637..937O,
2006ApJ...640..156G}.  In order to be detected with a high signal-to-noise
ratio, matched filtering algorithms sift through the LIGO data stream
looking for such circular inspirals \citep{2008PhRvD..77f2002A}, but might
miss many eccentric events
{\new \citep{PhysRevD.60.124008,2007arXiv0712.3199T,2008ApJ...681.1431M}}.
In principle, eccentric inspirals are well suited for detection as well, however they
have not been expected to be a significant source for terrestrial detectors
\citep[except see][]{2003ApJ...598..419W}.

In this paper, we propose an important additional source of
gravitational wave sources, which preferentially form binaries that
are still eccentric as they merge. In galactic nuclei with
supermassive black holes (SMBH) of mass $\msmbh < 10^7\,\Msun$,
relaxation times are often less than a Hubble time, and can result in
the formation of steep density cusp of stars and stellar mass black holes (BHs).  Indeed, as many
as $\sim 20,000$ BHs are expected to have
segregated into the inner $\approx 1\,$pc of the Milky Way
\citep[][hereafter HA06]{1993ApJ...408..496M,2000ApJ...545..847M,
  2006ApJ...649...91F,2006ApJ...645L.133H}.  When two BHs have a close
encounter, they can release sufficient amount of energy in GWs to form
a tight binary, and merge in less than a few hours.  These mergers are
almost always eccentric, in contrast to circular inspirals
expected in globular clusters. A similar process can occur in the
runaway growth of BHs when no SMBH is present that will also result in
eccentric mergers; however, such systems are inherently unstable and
not long-lived
\citep{1987ApJ...321..199Q,1989ApJ...343..725Q,1990ApJ...356..483Q,1993ApJ...418..147L}.

The GWs generated during the evolution of these orbits are endowed with a
rich structure.  For sufficiently small impact parameters, the GWs can be
detected with second generation GW instruments already during first passage
\citep{2006ApJ...648..411K}. Subsequently, the eccentricity is close to
unity for several thousand orbits after the formation of the binary,
generating a train of short-duration distinct GW bursts with a continuous
frequency spectrum.  As the eccentricity gradually decreases, the
separation between subsequent bursts decreases, and the waveform becomes
continuous in the time-domain and decouples into discrete harmonics in
frequency. Interestingly, as we show below, the eccentricity is
non-negligible all the way to coalescence.  The full waveform,
consisting of a correlated set of distinct GW bursts, which evolves into a
continuous eccentric inspiral signal, would not be discovered in the GW data
by existing data analysis techniques such as GW burst search algorithms,
nor by small eccentricity inspiral templates, since the contribution of
individual bursts to the GW power is small and the contribution of higher
harmonics is significant throughout the evolution
\citep[see][for a similar discussion for extreme mass ratio GW bursts during encounters with a supermassive black hole]{2008ApJ...675..604Y}.

Eccentric inspirals are more luminous in GWs than circular
inspirals and extend to higher frequencies, and so may be detectable to
higher redshifts.  Furthermore, these events may be capable of detecting
intermediate mass black holes using terrestrial instruments
\citep{2007PhRvL..99t1102B,2008ApJ...681.1431M}.  The evolution of the signal during the
initial phase of eccentric inspirals can be described using a small number
of parameters, allowing one to construct very efficient and sensitive data
analysis algorithms to search for these signals in the LIGO
data. Additionally, these signals spend a longer physical time in the LIGO
band, possibly reducing the false alarm rate for a fixed number of templates that
cover the observational period. The modulation introduced by the Earth's rotation
during such an event can be utilised to further improve the measurement
accuracy for the source position.

The GW waveforms have been studied for eccentric inspirals through several
different approaches. Post-Newtonian (PN) waveforms exist up to 1.5PN
order beyond the leading order gravitational radiation effects ($\propto v^3$)
for general mass ratios with spins
\citep{Vasuth:2007gx,2008PhysRevD.77.104005},
{\new 2PN without spins \citep[($\propto v^4$, including the effects of
eccentricity and radiation reaction]{2004PhRvD..70f4028D} and 3PN
including eccentricity but without spins and also neglecting radiation
reaction \citep[$\propto v^6$,][and references
therein]{2008PhRvD..77f4035A}, 5.5PN for extreme mass ratios without spins
\citep[$\propto v^{11}$,][]{1996PThPh..96.1087T}. The evolution of a
binary is chaotic at (or beyond) the 2PN approximation if the component
BHs have spins \citep{2006PhRvD..74l4027L,2007PhRvD..76l4004W} and also
possibly for nonspinning BHs approaching the unstable circular
orbit \citep[i.e. exhibiting zoom-whirl
orbits,]{2007CQGra..24...83P,2008arXiv0802.2520W}.}
Numerical relativity has mostly focused on the circular inspiral for
comparable mass BHs \citep[and references therein]{\baker,\berti,2008arXiv0804.4184B}, or extreme
mass ratios \citep{2004PhRvD..69d4025M,2007CQGra..24...83P,2007PhRvD..75b4005B}.
Recently, however, the waveforms have been evaluated for a handful of equal mass eccentric
initial conditions
\citep{2007arXiv0710.3823S,2008arXiv0802.2520W,2008arXiv0806.1037H,2008PhRvD..77h1502H}.

In this paper, we perform two separate analyses to two different
problems.  First, in \S~\ref{sec:massseg}, we determine the multi-mass
distribution of BHs in galactic nuclei for a variety of mass models.
In \S~\ref{sec:GB}, we determine the formation rate of
binaries due to gravitational wave capture in a variety of galactic
nuclei. In \S~\ref{sec:detection} we describe the general features
of the generated GW waveform, calculate the expected signal-to-noise ratio
for its detection with second generation terrestrial GW instruments and
determine the expected detection rate of such mergers.  We also show that
the eccentricity distribution of such sources will distinguish binaries
formed in the manner outlined in this paper from BH-BH binaries formed
dynamically in massive star clusters.

\section{Mass Segregation}
\label{sec:massseg}

\citet[hereafter BW76]{1976ApJ...209..214B} were the first to
correctly analyse the relaxation of a stellar population around a
central point mass, {\new although their analysis was highly idealised}. They
first derived the Fokker-Planck equations for a spherically symmetric
distribution of a single-mass population of low mass stars around a
massive black hole ($M_* \ll \msmbh$).  They found that after about
one half of a relaxation timescale the mass density profile of stars
reaches a steady-state and forms a power-law cusp with respect to
radius, $\propto r^{-\alpha}$, around the massive central object, with
$\alpha = 7/4$.  In a second paper, they extended their analysis to
look at a multi-mass system and included effects of the loss-cone
\citep{1976Natur.262..743S} that results from the disruption of stars
by the central black hole \citep[][hereafter
BW77]{1977ApJ...216..883B}.  They found that in a two-mass system the
more massive objects segregate from the lower mass stars by forming a
steeper power-law density profile than the stars.  This is in stark
contrast to the evolution of a two-mass cluster of stars without a
SMBH, which can eventually lead to the so-called Spitzer-instability
\citep{1969ApJ...158L.139S}, in which the high mass stars decouple
from the low mass objects entirely.

For a population of stars that are sufficiently old ($\ga 100\,\myr$),
stellar mass BHs with mass $\sim 10\,\Msun$ are expected to become the most
massive objects in the system and will begin to segregate from the main
sequence stars due to dynamical friction \citep{\morris}.  Looking at our
own Galactic Center and using a typical stellar mass function, \citet{\meg}
estimated that $\sim 2.5 \times 10^4\,$ BHs should have segregated into the
inner pc.  Indeed, HA06 followed the work of BW77 by solving the time
dependent Fokker-Planck equations and confirmed these results. They found
that $\sim 18,000$ BHs of $10\,\Msun$ each, should segregate to form a
steep density cusp around the SMBH in our galaxy, \sag, with $\alpha_{\rm
BH} \approx 2$, and that the stars and lighter compact objects form a
shallower cusp with $\alpha_* \approx 1.4$.

\citet[hereafter FAK06]{2006ApJ...649...91F} have so far provided the most
robust analysis of the segregation and distribution of a multi-mass system
of BHs and stars in the nuclei of galaxies \citep[for a review, see][]{2007astro.ph..3495A}.
The authors combined large $N\sim 10^6$ Monte-Carlo Fokker-Planck simulations with a population
synthesis code \citep{2002ApJ...572..407B} to simulate the segregation and
subsequent formation of a density cusp for an old population of stars.
Their simulations were in remarkably good agreement with previous analyses
\citep[e.g.][HA06]{\morris, \meg}.  In their simulations of a
Milky-Way--like galactic nucleus they found a similar overall number of
stellar mass BHs ($\sim 20,000$)
within 1$\,$pc of a $3.5\times10^6\Msun$ SMBH.  However, they found the BHs
follow a slightly shallower density profile than HA06, with $\alpha_{\rm
BH} \approx 1.8$.  Despite the level of detail in the simulations of FAK06,
the simulations still have their limitations.  Because of the large range
in scales needed to follow the mass segregation, and subsequently the large
number of particles in that volume, each star cannot be modelled by a
single particle.  The densest, most interesting regions of their
simulations still suffer from small number statistics, and can't reveal the
precise expectations for the mass distribution of stars and BHs. In order
to calculate the rate of gravitational wave capture events one needs to
apply a high-resolution method (like HA06) to multi-mass distribution
of BHs for calculating the rates and detectability of the events.

Although the BHs are expected to dominate relaxation and dynamical
encounters in the inner $0.1\,$pc around \sag, their presence has so
far eluded detection. X-ray observations of our own galaxy and others
may provide the first window to this interesting stellar population.
In our own galaxy, \citet{2005ApJ...622L.113M} have found an
overabundance of X-ray sources in the vicinity of \sag, which seems to
follow the underlying distribution of stars
\citep[][]{2006ApJS..165..173M}.  For large distances, the X-ray
sources appear to follow the BH distribution as well (FAK06).  In
other galaxies, many BHs may interact with the dense accretion disks
around SMBHs, and in some circumstances be intrinsically more luminous
than the SMBH \citep{2007MNRAS.377..897D, 2007MNRAS.377.1647N}.

The existence of a nuclear population of BHs may also be revealed
through dynamical encounters with observable stars. Indeed,
\citet{2005ApJ...622..878W} have shown that a decade of observations
with the next generation of $30\,$m class telescopes may detect the
scattering of the BHs with the so-called 'S-stars' that are within
$\approx 0.04\,$pc of \sag. Microlensing by the BHs is expected to
produce a much weaker signature
\citep{2000ApJ...545..847M,2001ApJ...551..223A, 2001ApJ...563..793C}.
Strong encounters between the BHs and the same population of S-stars
may dynamically eject the stars with high enough velocity to escape
the potential of the galaxy \citep{2008MNRAS.383...86O} and populate
the halo with the observed hypervelocity stars
\citep{2005ApJ...622L..33B,2006ApJ...640L..35B}.  Strong encounters
are also expected to tidally spin up long-lived stars, and so the BH
population may be inferred through spectroscopic measurements of the
spin of lower-mass stars \citep{2001ApJ...549..948A}.  {\new
  Alternatively, one may infer the presence of the BHs by looking at
  density distribution of old pulsars \citep{2002ApJ...571..320C}.}

Finally, future GW observations with the space based GW interferometer
{\em LISA}\footnote{http://www.lisa-science.org/} or the Earth-based
second generation GW instruments are expected to detect the inspiral
of such BHs into SMBHs
\citep{2004CQGra..21S1595G,2006ApJ...649L..25R,2007MNRAS.378..129H,2008ApJ...675..604Y}
or IMBHs \citep{2007PhRvL..99t1102B,2008ApJ...681.1431M},
respectively, several times per year.  Given the test particle limit
of the BH, such encounters are expected to be a stringent test of
General Relativity in the strong-field regime
\citep{2004PhRvD..69l4022C,2006PhRvD..74b4006A}.

An interesting method for detecting a dense population of BHs is
through the GWs they produce as they scatter on each other and get
captured into a tight binary. \citet{2006ApJ...648..411K} estimated
the expected detection rates of the first hyperbolic passage between
such stellar mass black holes in globular clusters, and found rates to
be typically less than $0.1{\,\rm yr}^{-1}$ for second generation GW
instruments. Here, we consider the detectability of the much stronger
GWs that follow once the eccentric binary has formed in galactic
nuclei in this way and evolves toward coalescence.  We explore this
new method in \S~\ref{sec:GB} and~\ref{sec:detection}. However, in
order to completely determine this rate we need to know how a
multi-mass distribution of BHs relaxes around a SMBH in galactic
nuclei.  In the remainder of this section we calculate the
steady-state distribution of BHs for a variety of models.

\subsection{Fokker-Planck Equations}
In our calculations, we follow the analysis of BW77, who derived the
multi-mass Fokker-Planck equations for a spherically symmetric and
isotropic distribution of stars around a SMBH, $f_M(r,v) = f_M(E)$,
where $E=(1/2) M v^2 - G M \msmbh/r$ is the mechanical energy of bound
orbits in the galactic cusp that are outside the loss cone (see below).
Much of our analysis follows that of HA06, whose
terminology we use throughout our calculations.  The main difference
in our work is that we look at a multi-mass distribution of BHs with
slightly different initial conditions.

The SMBH determines the motion and dynamics of stars within its radius
of influence, $r_i = G \msmbh / \sigma_*^2$, where $\msmbh$ is the
mass of the SMBH, and $\sigma_*^2$ is the one-dimensional velocity
dispersion of the underlying stellar population in the nucleus near
the SMBH. For our calculations, we define a dimensionless time
variable $\tau = t/t_r$, where $t_r$ is the relaxation timescale at
the radius of influence,
\begin{equation}
  \label{eq:relaxtime} t_r = \frac{3(2\pi \beta_* /M_*)^{3/2}}{32 \pi^2 G^2
 M_*^2 n_* \ln{\Lambda}},
\end{equation}
where $n_*$ is the number density of stars at $r_i$ for the dominate
population of stars of mass $M_*$, $\beta_* = M_* \sigma_*^2$, and $\ln
\Lambda \approx \ln (\msmbh/M_*)$ is the Coulomb logarithm. We also define
a dimensionless unit of energy, $x = -(M_*/M)(E/\beta_*)$, and a
dimensionless distribution function $g_M(x) = [(2 \pi \beta/M_*)^{3/2}
n_*^{-1}]f_M(E)$. In these units, the Fokker-Planck equation is (BW77,HA06)
\begin{equation}
  \label{eq:fokkerplanck}
  \frac{\partial g_M(x,\tau)}{\partial \tau} = -x^{5/2}
  \frac{\partial}{\partial x} Q_M(x) - R_M(x),
\end{equation}
where
\begin{eqnarray}
  \label{eq:flowrate}
  Q_M(x) = \sum_{m} \frac{M}{M_*} \frac{m}{M_*} \int_{-\infinity}^{x_D}
  \rmd x' [{\max} (x,x')]^{-3/2} \nonumber \\
  \times \left(g_M(x) \frac{\partial
      g_{m}(x')}{\partial x'} - \frac{m}{M}g_{m}(x')\frac{\partial
      g_M(x)}{\partial x}\right).
\end{eqnarray}
is the rate stars of mass $M$ flow to energies larger than $x$, and
$R_M(x)$ is the rate stars are destroyed by the SMBH.  We are
interested in the distribution of BHs in the inner pc and therefore
look only at the empty-loss cone regime.  Thus, the angular momentum
averaged destruction rate of stars of mass $M$ with energy $x$ is (HA06)
\begin{equation}
  \label{eq:losscone}
  R_M(x) = \frac{g_M(x)}{\tau_r(x) \ln[J_c(x)/J_{\rm LC}]},
\end{equation}
 where $\tau_r$ is the approximate
local relaxation time at radius $r = r_i/(2 x)$,
\begin{equation}
  \label{eq:taur}
  \tau_r = \frac{M_*^2}{\sum_{M} g_{M}(x) M^2},
\end{equation}
$J_c(x) = G \msmbh \sigma_*^{-1} (2 x)^{-1/2}$
is the angular momentum of a
circular orbit at a radius $r = r_i / (2 x)$, and $J_{\rm LC}$ is the
maximum angular momentum for the object to be destroyed or consumed by
the SMBH.  For stars
$J_{\rm LC} = G \msmbh \sigma_*^{-1} (2 (x + x_{\rm td}))^{1/2} x_{\rm td}^{-1}$
corresponds to a star coming within tidal
disruption radius of the SMBH, where
$x_{\rm td} = (\msmbh/M_*)^{-1/3} r_i/R_*$
and $R_*$ is the radius of the star.  BHs plunge into
the SMBH when $J < J_{\rm LC} = 4 G \msmbh/c$.

We numerically integrate Eq.~(\ref{eq:fokkerplanck}) with our
initial conditions given in \S~\ref{initconditions} until $g_M(x)$
reaches a steady state for all $M$.
For all of our runs this occurs before $\tau = 1$.
A computationally straightforward simple iterative way to generate the
steady state solution for Eq.~(\ref{eq:fokkerplanck})
satisfying $\frac{\partial g_M}{\partial \tau}\equiv 0$
is to follow the time evolution of the distribution from the initial
conditions.

After the integration is complete, we calculate the number density of
stars at radius $r$,
\begin{equation}
  \label{eq:numdens}
  n_M(r) = 2 \pi^{-3/2} n_* \int_{-\infinity}^{\phi(r)}\rmd x g_M(x) (\phi(r)-x)^{1/2},
\end{equation}
where {\new $\phi(r)$} is the dimensionless specific
potential.  Until now, we have assumed that the stars were on entirely
Keplerian orbits around the SMBH (i.e. that $\phi(r) = r_i/r$).  This
assumption was necessary in order to simplify the Fokker-Planck
equations to Eq.~(\ref{eq:fokkerplanck}) (BW76).  However, for our
calculations in \S~\ref{sec:GB}, it is important that the BH density
goes to zero as $r$ approaches infinity.  {\new We therefore calculate
$\phi(r)$ and $n_M(r)$ iteratively after calculating the steady-state
distribution functions $g_M(x)$. We do this by sequentially solving for $n_M(r)$
(Eq.~\ref{eq:numdens}) and
\begin{equation}
  \label{eq:phi}
  \phi(r) = \frac{r_i}{ r} - \int_0^r \frac{M(<r) r_i}{
    \msmbh r} \rmd r
\end{equation}
where $M(<r) = \sum_M \int_0^r 4 \pi r^2 n_M(r) \rmd r$ is the total
mass of stars interior to $r$, thus accounting for the stars'
contribution to the potential as well as the SMBH's contribution. We
find that we converge to a single solution for $\phi(r)$ and $n_M(r)$
before about 6 iterations.  This solution to the number density and
potential is fully consistent with the distribution functions
$g_M(x)$.  However, Equations~\ref{eq:fokkerplanck} \&
\ref{eq:flowrate} implicitly assume that, as $r \rightarrow \infty$,
the potential approaches zero and the number density approaches a
constant.  This assumption is approximately followed as long as the
core radius of the isothermal sphere $r_c = [9\sigma_*^2/(4\pi G
\rho_0)]^{1/2}$ \citep{1987gady.book.....B} is larger than the radius
of influence of the SMBH.} We also note that the number density is the
statistical average. In cases of the small numbers of BHs, we assume
that values we determine will match the average over many
galaxies. However, we caution that the dynamics of such systems will
likely behave differently than assumed here.

\subsection{Initial Conditions and BH Mass Distributions}
\label{initconditions}
The initial conditions of our models are described entirely by the
distribution functions of unbound stars $g_M(x<0)$. 
Sufficiently far from
the radius of influence of the SMBH the population of visible stars appear
to be similar to an isothermal sphere, with a constant velocity dispersion
and a number density distribution that scales as 
$\propto r^{-1.8}$
\citep{2003ApJ...594..812G,2007AA...469..125S}. 
We therefore set $g_*(x) = \exp(x)$ for $x<0$, {\new which corresponds to an isothermal sphere of stars}.  
We look at only one population of stars of mass $M_* =
1\,\Msun$.  This is justified by previous analyses of mass segregation that
show that the underlying stellar population, being considerably less
massive than BHs, tends to follow the same distribution function
independent of mass where the BHs dominate the relaxation (HA06, FAK06).
We confirm this conclusion in \S~\ref{sec:massseg:results}.  Although there
is a significant fraction of more massive stars in nuclei, their lifetimes
are expected to be less than the relaxation timescale at most $r$ and {\new are in such small number} that
they should not contribute to the dynamics significantly.

We assume that the BHs initially follow the distribution of the stars, and
amount to a total fraction, $C_{\rm BH} = \sum_M C_M$, of the number
density.  BH natal kick velocities are expected to be much less than
the velocity dispersions of the systems we are interested in
\citep{1996ApJ...473L..25W,2005ApJ...625..324W}, and so we expect the
BHs' velocity dispersion to be similar to the stars'.  Therefore, we
set $g_M(x) = C_M \exp(x)$,
yielding a BH number density of $C_M n_*$ near $r_i$. This is in
contrast to the work of BW77 and HA06a, who both assume that the
population of stars was already in thermal equipartition (i.e. $M
\beta_M = M_* \beta_*$).  BW77 addressed the distribution of stars in
globular clusters, which have half-mass relaxation times much shorter
than the Hubble time and so the entire system should reach
equipartition.  In contrast, in galactic nuclei the relaxation
timescale for large radii is usually longer than the age of the
universe, and so the source population could not reach a complete
equilibrium. HA06 made a numerical error in how they normalised the BH
distribution and so had reasonable agreement between their work and
previous works.  However, with the proper normalisation, they would
have found the number of BHs to be $(M / M_*)^{3/2} = 10^{3/2}$ times
larger than their results indicated.  This is because in the case of a
velocity dispersion independent of mass, the ratio of number densities
of objects is described by $n_M/n_* = C_M/C_*$, whereas for a
population in thermal equilibrium we get
$n_M/n_* = C_M M^{3/2}/(C_* M_*^{3/2})$.
Because the steady-state distribution function is most sensitive to
$g(0)$ and not the functional form $g(x<0)$, their results are similar
to ours when one uses the incorrect normalisation.

One goal of our analysis is to see if the mass distribution of BHs may be
determined either through dynamical interactions with their environment, or
through the release of GWs.  We therefore look at a variety of
time-independent BH mass functions parametrised as power-laws.  The models
are determined by four basic parameters: the total number fraction of BHs,
$C_{\rm BH} = \sum_M C_M$; the slope of the BH mass function, $\beta$,
where $\rmd n_M / \rmd M \propto M^{-\beta}$; the minimum BH mass, $M_{\rm
min}$; and, finally, the maximum BH mass in the nucleus, $M_{\max}$. In
our calculations, we approximate the distribution of the BHs with $N=9$
discrete distribution functions with
\begin{equation}
  \label{eq:bhdist}
  C_i = \int_{M_{\rm min}+i \Delta M}^{M_{\rm min}+(i+1) \Delta M}
  \gamma M^{-\beta} \rmd M,
\end{equation}
where $\Delta M = (M_{\max} - M_{\rm min})/N$, and $\gamma$ is
normalised such that $\sum_i C_i = C_{\rm BH}$.  We set the mass for
each distribution function to be the average mass for that bin,
\begin{equation}
  \label{eq:bhmass}
  M_i = \int_{M_{\rm min}+i \Delta M}^{M_{\rm min}+(i+1) \Delta M}
  (\gamma / C_i) M^{-\beta+1} \rmd M.
\end{equation}
We describe the numbers used above in detail below, and outline them in
Table~\ref{table1}.

\begin{table*}
  \centering
  \begin{minipage}{140mm}
    \caption{\label{table1} BH Models and Results. {\new Unless otherwise
      noted, these results apply to a Milky Way like galaxy with
      $\msmbh = 3.5\times 10^6\,\msun$, $\sigma_* = 75\,\kms$ and $r_i
      = G \msmbh/\sigma_*^2 = 2.7\,$pc.} The columns are, from left to
      right, (1) the model name, (2) the slope of the BH mass
      function, (3) the minimum BH mass, (4) the maximum BH mass, (5)
      the fraction of stars that are BHs, (6) the merger rate of GW
      capture binaries per model galaxy, and (7) the expected AdLIGO
      detection rate as calculated in \S~\ref{sec:detection}.}
    \begin{tabular}{@{}lcccccc@{}}
      \hline
Model        & $\beta$ & $M_{\rm min}$ & $M_{\max}$ & $n_{\rm BH}/n_*$  & Merger Rate per Galaxy  & AdLIGO Detection Rate   \\
             &         &  ($\Msun$)   &    ($\Msun$) &                  &       (yr$^{-1}$)          &  ($\xi_{30}\yr^{-1}$)            \\
\hline
A           &   2     &     5        &     10       &       0.001       &    $2.2\times 10^{-10} $   &     5.8                 \\
A$\beta3$   &   3     &     5        &     10       &       0.001       &    $2.0\times 10^{-10} $   &   5.2                   \\
B           &   2     &     5        &     15       &       0.001       &    $2.8\times 10^{-10} $   &   16                   \\
B$\beta3$   &   3     &     5        &     15       &       0.001       &    $2.5\times 10^{-10} $   &   13                  \\
B-1         &   2     &     5        &     15       &       0.1       &    $5.3\times 10^{-9} $   &   220                   \\
B-2         &   2     &     5        &     15       &       0.01       &    $8.3\times 10^{-10} $   &   40                   \\
Be5\footnote{This is evaluated for a SMBH with $\msmbh = 1\times 10^5\,\Msun$ and $\sigma_* =30\,\kms$.}    &   2     &     5        &     15       &       0.001       &    $2.8\times 10^{-10} $   &   15                   \\
C           &   2     &     5        &     25       &       0.001       &    $3.8\times 10^{-10} $   &   46                  \\
D           &   2     &     10        &     15       &       0.001       &    $3.4\times 10^{-10} $   &   23                   \\
E           &   2     &     10        &     25       &       0.001       &    $3.2\times 10^{-10} $   &   12                   \\
E-1         &   2     &     10        &     25       &       0.1       &    $1.2\times 10^{-8} $   &   1200                   \\
E-2         &   2     &     10        &     25       &       0.01       &    $1.3\times 10^{-9} $   &   150                   \\
F-1         &   2     &     10        &     45       &       0.1       &    $1.5\times 10^{-8} $   &   2700                   \\
BSR\footnote{There is an additional population of $25\,\Msun$ BHs with  $n_{\rm bh}/n_* = 10^{-5}$}   &   0     &     5       &     10      &       0.001         &    $2.9\times 10^{-10} $   &   43               \\
\hline
\end{tabular}
\end{minipage}
\end{table*}

The number fraction of BHs in galactic nuclei is sensitive to the
initial mass function (IMF) of high mass stars.  For a Kroupa IMF
{\new \citep{2003ApJ...598.1076K}}
we estimate $C_{\rm BH} \approx 0.001$ by assuming that all stars with mass
$M>20\,\Msun$ become BHs.  A similar distribution of stars formed
uniformly throughout time was found to be consistent with the K-band
luminosity distribution of stars in the galactic centre
\citep{1999ApJ...520..137A}.  However, more recent observations of our
Galactic Center suggest the IMF slope of high mass stars may be
considerably shallower than those in regular clusters \citep[see,
e.g.,][]{2005MNRAS.364L..23N, 2006MNRAS.366.1410N,2006ApJ...643.1011P,
  2007ApJ...669.1024M}.  \citet{2006ApJ...643.1011P} found that the
stars in an apparent disk around Sgr A* are best fit with a flattened
IMF $\propto m^{-\beta_{\rm imf}}$ with $\beta_{\rm imf} \approx
0.85-1.35$ and a depletion of low mass stars.  This is consistent with
the work of \citet{2005MNRAS.364L..23N}, who constrained the number of
young low mass objects with X-ray observations, and determined there
to be a deficiency of stars with mass $\lesssim 3\,\Msun$.  Most
recently, \citet{2007ApJ...669.1024M} looked at late-type giants and
concluded that the MF of high mass stars may be as shallow as
$\beta_{\rm imf} \approx 0.85$.  A shallow MF is also expected
theoretically \citep{\morris} and is consistent with the most recent
hydrodynamic simulations of star formation in the
Galactic Center
\citep{2008ApJ...674..927A}. Considering the abundant evidence for an
alternative IMF in galactic nuclei, we therefore look at a wide range
of number fractions, setting $C_{\rm BH} = 0.001, 0.01,$ and $0.1$,
which roughly corresponds to $\beta_{\rm imf} = 2.3, 1.5,$ and $0.8$
respectively.

The mass distribution of BHs is relatively unconstrained.  Currently,
the best constraints come from the couple of tens of X-ray binaries
with dynamically determined BH masses \citep{2006ARA&A..44...49R}.
Each measurement often has considerable uncertainty, but the masses
seem to span a range of $\sim 3 - 18\,\Msun$. There is now strong
dynamical evidence for BHs with even greater masses, up to
$23$--$34\,\Msun$,
in low metallicity environments
\citep{2007Natur.449..872O,2007ApJ...669L..21P,2008ApJ...678L..17S}. Nevertheless,
these observations suffer from a severe observational bias; the BHs
must be in a close binary to be observed.  Theoretical estimates for
the mass distribution are also highly uncertain.
\citet{2004ApJ...611.1068B} have used their sophisticated population
synthesis models to determine the expected distribution of BH masses
in a variety of environments.  Typically, they expect most BHs in high
metallicity environments to have a uniform distribution of masses
between $\sim 5 - 10\,\Msun$, but do find that significantly more
massive BHs may form from the merger of the BH with a high mass
companion star. Unfortunately, these massive BHs would not be found in
X-ray binaries, unless they were introduced into one dynamically.
Given these uncertainties, we choose to parametrise the mass
distribution of BHs as a power-law $dn_M/dm \propto m^{-\beta}$, where
we consider $\beta= 2$ and $\beta = 3$.  We look at the importance of
the highest mass BHs, by modelling different upper limits on BH mass,
$M_{\max}$. We also have a model based on the work of
\citet{2004ApJ...611.1068B}. In Model BSR, we use a flat $\beta = 0$
model of BHs with masses between $5$ and $10\,\Msun$ with a fraction
$\sim 0.01$ of BHs with mass $25\,\Msun$.  This is consistent with
their Model C2 of solar metallicity stars and based on their Figure~7.

For the remainder of \S~\ref{sec:massseg}, we focus our results on
Milky Way like nuclei, and assume that $\msmbh = 3.5\times
10^6\,\Msun$ \citep{2005ApJ...620..744G,2005ApJ...628..246E} and that
the velocity dispersion of the stars and stellar BHs is $\sigma_* =
75\,\kms$ (HA06). In \S~\ref{sec:GB} we also
consider other galaxies that have relaxation times short enough to
reach steady state in a Hubble time with $\msmbh \sim
10^4-10^7\,\Msun$, in order to calculate the overall rate of GWs
sources in the universe.

\subsection{Results and Implications}
\label{sec:massseg:results}

\begin{figure*}
  \centering
  \includegraphics[width=\columnwidth]{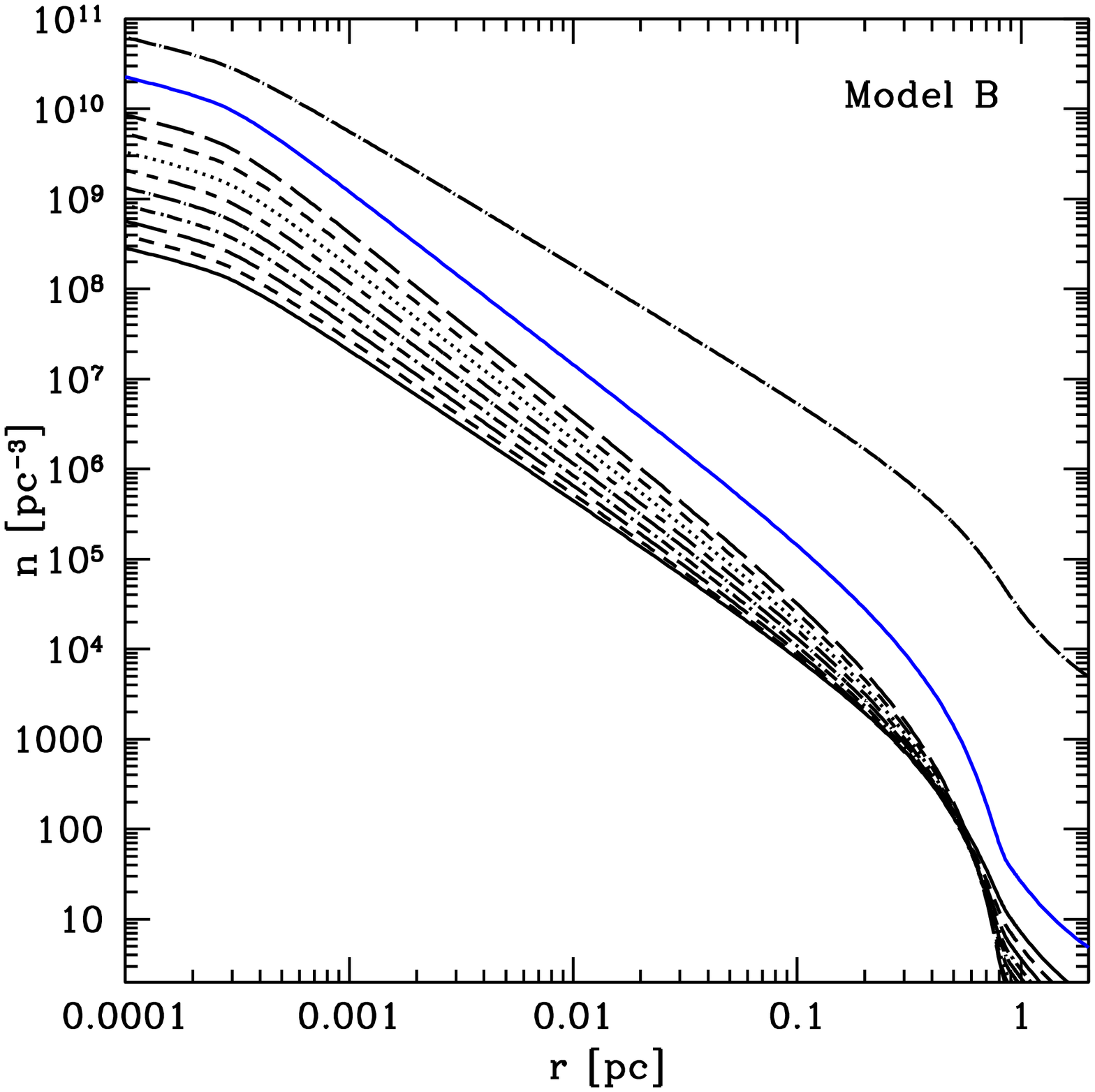}
  \includegraphics[width=\columnwidth]{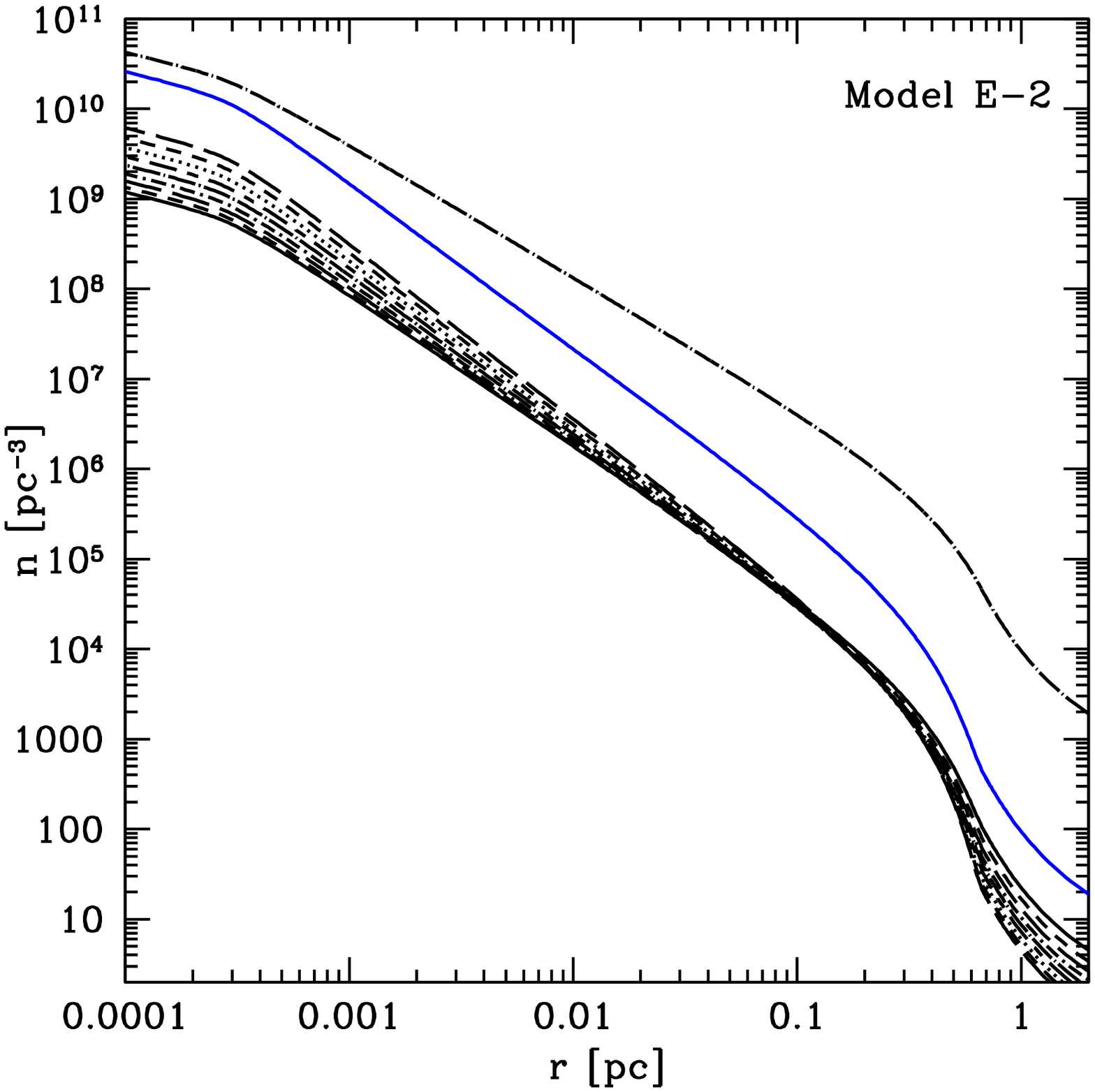}
  \caption{\label{fig:numberdensity} The number density of stars and
    BHs for Model B (left) and E-2 (right). The top dash--dotted line is the
    number density of the stars as a function of radius. The
    alternating dotted and dash-dotted lines show the number density
    of the separate mass bins used in our calculations with the most
    massive BHs having the largest number density interior to $r
    \approx 0.3-0.6\,$pc. The solid blue line is  the total BH number
    density $=\sum_M n_M(r)$.  Near $r \approx 1\,$pc the potential of
    the stars and BHs is equal to the Keplerian potential of the SMBH.
    The rapid drop in the number of BHs drops rapidly here because of
    the rigid boundary condition at $\phi = 0$. In all of our
    simulations, the bin of the most massive BHs dominates the number
    density of BHs in the inner $\sim 0.1\,$pc. }
\end{figure*}

For all of our models, independent of the BH mass function (namely,
$\beta$, and the mass range of BHs), the most massive BHs always become the
dominant BH species in the inner $\sim 0.1\,$pc of the galactic nucleus. In
Figure~\ref{fig:numberdensity}, we plot the number density of the stars and
BHs as a function of radius.  As has been found previously (BW77, HA06,
FAK06), deep in the Keplerian potential of the SMBH, the distribution
functions of the stars and BHs become power-laws of the negative specific
energy of the objects $g_m(x) \propto x^{p_m}$, and hence have a power-law
density profile $\propto r^{-3/2-p_m}$.  We find that throughout all of our
simulations, the exponent $p_m$ is best fit by a linear relationship
between the mass ratio of the object, $m$, and the most massive BH, $M_{\rm
max}$,
\begin{equation}
  \label{eq:powerlawrelation}
  p_m = p_0 \frac{m}{M_{\max}},
\end{equation}
where $p_0 \approx 0.5-0.6$.  We have found that $p_0$ usually has a
small scatter ($\sim 20\%$), depending on $M_{\max}$ and $\beta$;
however, given $p_0$, Eq.~(\ref{eq:powerlawrelation}) is often
accurate to $\lesssim 1\%$. This relationship is similar to that found
in BW77, who found that $p_0 \approx 0.25-0.3$, when they looked at
two different components with comparable number densities.
\citet{2008arXiv0808.3150A} attributes the steeper density profiles to
``strong'' mass segregation, where the relaxation of the system is
determined by the many low mass objects.  In their calculations, $p_0$
increases monotonically with $M_{\max}$, however the rate at which
it increases is rather slow after $p_0 \sim 0.5$.  In contrast, in our
simulations with a mass spectrum of massive objects and accounting for
the loss-cone, we do not find that the maximum power-law index $p_0$
to depend sensitively on the number or mass of the highest mass
objects.  However, we have not looked at the large range of parameters
of \citet{2008arXiv0808.3150A}.

\begin{figure*}
  \centering
  \includegraphics[width=\columnwidth]{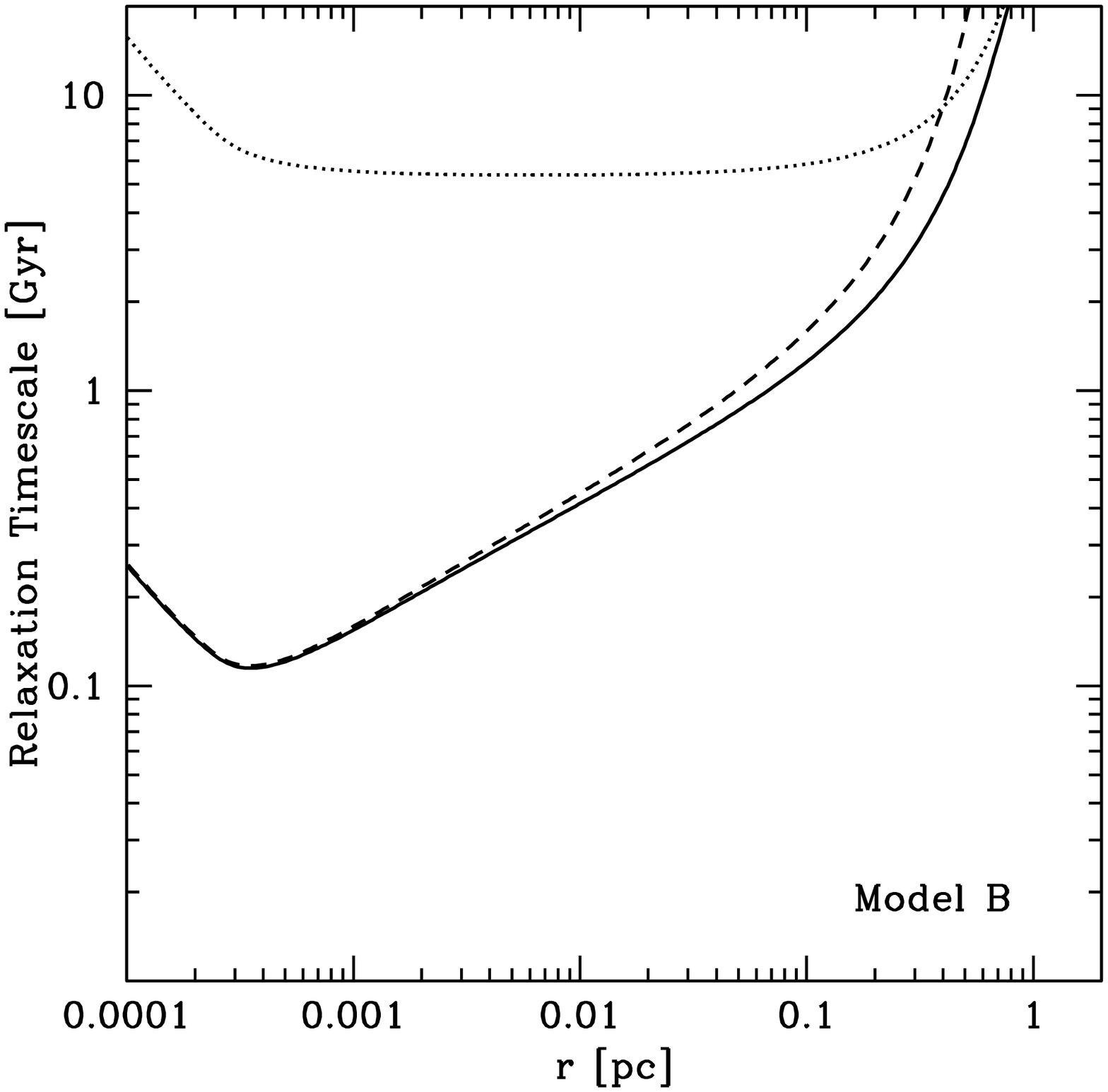}
  \includegraphics[width=\columnwidth]{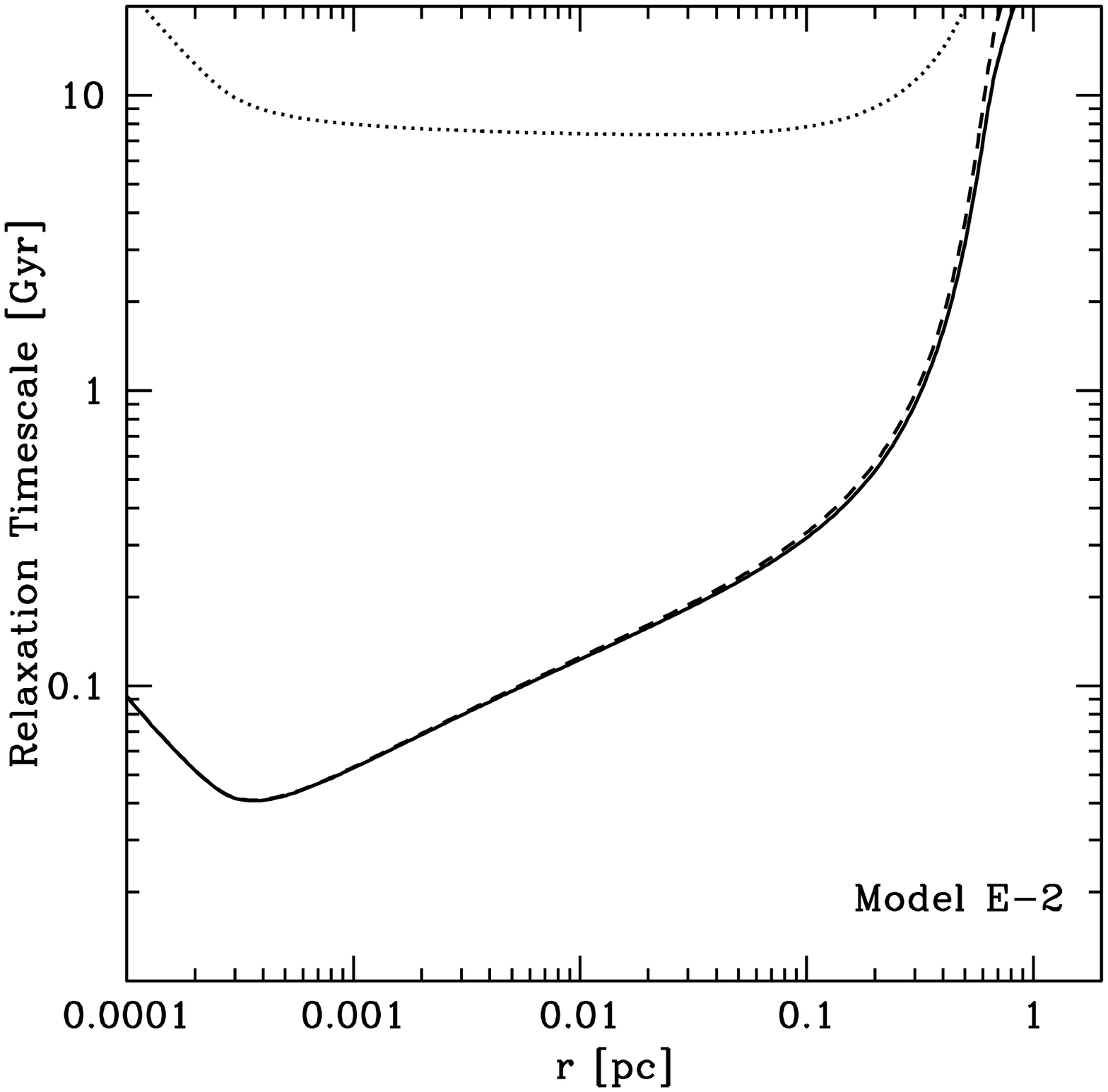}
  \caption{\label{fig:relax} The relaxation timescale as a function of
    radius for Model B (left) and E-2 (right) plotted in Gyr. The
    solid line shows the relaxation timescale for the entire system.
    The long dashed and dotted lines show the contribution of
    the BHs and stars respectively. Although the stars contribute to
    the formation of the cusp and are the most common objects in the
    system, their low mass precludes them from dominating the
    relaxation in the inner $\approx 0.6\,$pc of the cusp.  The break
    in the relaxation time at $\sim 0.0002\,$pc is due to our boundary
    condition that no stars have $x > 10^4$.  }
\end{figure*}

Mass segregation in galactic nuclei ceases when the BHs begin to dominate
the relaxation process in the inner cusp.  Typically, this is expected when
$n_{\rm BH} M_{\rm BH}^2 > n_* M_*^2$.  One therefore may try to detect the
presence of a cluster of BHs through their interactions with the luminous
stars in its vicinity.  We can estimate the relaxation timescale at a
radius $r$ from Eqs.~(\ref{eq:relaxtime})~\&~(\ref{eq:taur}), in terms of
the Keplerian potential of the SMBH
\begin{equation}
  \label{eq:relaxtimecusp}
  t_r(r) = \frac{3 (2 \pi v_c^2(r))^{3/2}}{32 \pi^2 G^2 \ln{\Lambda}
    \sum_M{M_M^2 n_M(r)}}
\end{equation}
where $v_c(r) = \sqrt{G \msmbh / r}$ is the circular velocity around
the SMBH at radius $r$.  In Figure~\ref{fig:relax}, we have plotted
the relaxation timescale as a function of radius for Models B and E-2.
The BHs begin to dominate the local relaxation processes at $r \sim
0.5\,$pc, consistent with the results of FAK06 and HA06.  We find that
the relaxation timescale in the galactic centre is highly sensitive to
the assumed mass distribution of the BHs, since the lower density of
the BHs is easily outweighed by their greater mass.  By finding a
tracer of the relaxation time in the inner $0.1\,$pc of the galactic
centre, it may be possible to determine if such heavy BHs even exist
in our galaxy \citep{2002ApJ...571..320C}.

As discussed above, a population of massive BHs in the galactic centre
may be revealed through strong encounters with their neighbouring
stars.  Because the BHs are more massive than the typical star, they
tend to drive stars out of the nucleus, creating a shallow stellar density
profile  (FAK06; HA06; for observations see
\citealt{2003ApJ...594..812G} and more recently
\citealt{2007AA...469..125S}).  \citet{\meg} proposed that a
population of such BHs may be seen through stars on very radial orbits
that are ejected by close encounters by the BHs.  In some extreme
encounters, \citet{2008MNRAS.383...86O} showed that strong
interactions between BHs and stars can lead to the stars being ejected
from the Milky Way altogether as the observed hypervelocity stars
\citep{2005ApJ...622L..33B,2006ApJ...640L..35B}. The results of
\citet{2008MNRAS.383...86O} were sensitive to the assumed BH mass
distribution.  Self-segregation of the BHs is an important
consideration when evaluating the likelihood of the scenario, and the
velocities of ejected HVSs \citep{2007MNRAS.379L..45S}.

As a more extreme example of a strong encounter, BHs and stars may
physically collide into each other \citep[][FAK06]{\morris} producing
strong X-ray and UV flares in galactic nuclei.  Interestingly, we find
that the total rate of physical collisions between stars and BHs can
be comparable to the rate stars are tidally disrupted by the
SMBH, and may be visible in variability surveys of galactic nuclei.
Such encounters may have smaller bolometric luminosities than stars
disrupted by the SMBH due to their lower Eddington luminosity.

\section{Binary Formation, Inspiral, and Event Rates}
\label{sec:GB}

When two compact objects have a close encounter, they can emit
sufficient energy through GWs that they become bound and form a
binary.  {\new To see the importance of such a process in galactic
  nuclei, we first estimate the rate of binary formation for two
  typical $10\,\msun$ BHs.  We present more detailed calculations
  later in this section. To first order, two BHs with an initial
  relative velocity $w$ will form a 
  bound binary due to the emission of GWs
  if they come within a distance
  $\approx 7.4\times 10^{-3}\,\rsun (w/100\,\kms)^{-4/7}$ at closest
  approach (see Eq.~[\ref{eq:rpmax}] below).  This corresponds to an initial impact parameter (the
  distance of closest approach if the BHs moved in straight lines) of (Eq.~[\ref{eq:bmax}])
  $b \la 2.4\,\rsun (w/100\,\kms)^{-9/7}$. We expect that a single BH
  will undergo such an encounter approximately every $(\pi b^2 n_{\rm
    BH} w)^{-1} = 3.4\times 10^{12} (n_{\rm BH}/10^6 {\rm
    pc}^{-3})^{-1} (w/100\,\kms)^{11/7}\,\yr$. The total rate of
  binary formation is then approximately (see, e.g., Eq.~[\ref{eq:diffsimpfin}]) $\int n_{\rm BH}^2 \pi
  b^2 w 4 \pi r^2 \rmd r$, where $w(r) \sim v_c(r) = \sqrt{G \msmbh/r}$
  is the Keplerian circular velocity at radius $r$.  Integrating from
  $r = 10^{-3}\,$pc to $r = 1 \,$pc, yields a rate of $3.4\times
  10^{-10}\,\yr^{-1} [N_{\rm BH}/ (2\times 10^4)]^2$ for $N_{\rm BH}$ BHs in
  a density cusp around the SMBH with $n_{\rm BH} \propto
  r^{-2}$. This is in good agreement with our more detailed
  calculations presented in \S~\ref{sec:rates}.

  In this section, we calculate the formation rate of such
  gravitational wave capture binaries in greater detail. {\new We
    first calculate the conditions necessary to form gravitational wave
    capture binaries and the properties of their eventual inspiral in
    different environments. We then calculate the rate of binary
    formation for the multi-mass, segregated clusters of BHs that we
    modelled in \S~\ref{sec:massseg}.  Finally, we extrapolate our
    results for galactic models, including the distribution of SMBHs
    in the universe and accounting for the statistical variance in the
    number density of stars surrounding the SMBHs, in order to estimate the
    comoving rate density of binary formation and inspiral.  }
 
In our calculations, we express our equations with the symmetric mass
ratio of the encounter $\eta = (m M)/ (m+M)^2$, and total mass of the
BHs $\mtot$.  With these variables, the reduced mass is $\mu = \mtot
\eta$.

}

\subsection{Orbital Evolution}

In the following sections, we shall work with dimensionless parameters
in units ${G} = {c} = 1$, where the total mass $\mtot$ has length and time
dimensions. We define the dimensionless pericenter distance as
$\rho_{p}=r_p/\mtot$, dimensionless semimajor axis for an eccentric orbit
with eccentricity $e$ as $\alpha\equiv a/\mtot = \rho_p/(1-e)$.
In these units, the mean orbital angular frequency from Kepler's law is
simply $\omega_{\rm orb}=\mtot^{-1} \alpha^{-3/2}=\rho_p^{-3/2}(
1-e)^{3/2}$, the angular frequency at pericenter passage is
$\omega_{p}=\mtot^{-1} \rho_p^{-3/2}(1 +e)^{1/2}$, the orbital time is
$\Delta t_{\rm orb}\equiv f_{\rm orb}^{-1}\equiv 2\pi\omega_{\rm
  orb}^{-1} $. We define the characteristic time of pericenter passage
as $\Delta t_{p}\equiv f_{\rm p}^{- 1}\equiv 2\pi\omega_{p}^{-1}$,
which satisfies $\Delta t_{p}=\Delta t_{\rm orb}$ for circular
encounters, but is much smaller for eccentric encounters.

\subsubsection{First Passage, Formation of Binaries}

\label{sec:binform}
Two BHs of mass $m$ and $M$ will form a binary if they undergo a close
encounter and release enough energy to become bound, $\delta E > \eta \mtot w^2/2$, where
$w=|\bm{v}_m-\bm{v}_M|$ is the magnitude of relative velocity of the BHs at infinity,
and we ignore the presence of the SMBH during the encounter and assume that the
interaction is local. Given the relativistic nature of such events,
and the comparatively low velocity dispersions in nuclei, the
encounters are always nearly parabolic
\citep{1987ApJ...321..199Q,1993ApJ...418..147L}.  In this limit, the
amount of energy released during the encounter is
\citep{1963PhRv..131..435P, 1977ApJ...216..610T}
\begin{eqnarray}
\label{eq:de}
\delta E
&\approx&
-\frac{85\pi}{12\sqrt{2}} \frac{\eta^2\mtot^{9/2}}{r_{p}^{7/2}},
\end{eqnarray}
where $r_{p}$ is the distance of closest approach,
\begin{eqnarray}
r_{p} &=& \left(\sqrt{\frac{1}{b^2}+\frac{\mtot^2}{b^4 w^4}} +
  \frac{ \mtot}{b^2 w^2}\right)^{-1}\nonumber\\
&\approx& \frac{b^2 w^2}{2 \mtot}\left( 1 - \frac{b^2 w^4}{4\mtot^2}\right)
\label{eq:rmin}
\end{eqnarray}
and $b$ is the impact parameter of the encounter, and in the second line we have expanded to first order in $w/c$.

The final properties of the system are determined by the system's final energy,
$E_{\rm final}= \mtot \eta w^2/2+ \delta E$, and angular momentum,
$L_{\rm final} =  \mtot \eta b w + \delta L$, where
\begin{eqnarray}
  \label{eq:deltal}
   \delta L &\approx& -\frac{6\pi \mtot^4\eta^2}{r_p^2},
\end{eqnarray}
is the amount of angular momentum lost in GWs
\citep{1964PhRv..136.1224P}.  For nearly all of the encounters we find
$|\delta L| \ll \mtot \eta b w $ and so we set $\delta L = 0$.  If the total
energy of the system is negative, $E_{\rm final} < 0$, then the system
will remain bound with a semi-major axis,
\begin{equation}
  \label{eq:a}
  a_0 = - \frac{\mtot^2\eta}{2 E_{\rm final}},
\end{equation}
eccentricity,
\begin{equation}
  \label{eq:e}
  e_0 = \sqrt{1 + 2 \frac{E_{\rm final} b^2 w^2}{\mtot^3 \eta}},
\end{equation}
and pericenter, $r_{p0} = a_0 (1-e_0)$.

With the criterion $E_{\rm final} < 0$ for binary formation, the
maximum impact parameter for two BHs with relative velocity $w$ to
become bound is
\begin{eqnarray}
\label{eq:bmax}
b_{\max} &=& \left(\frac{340\pi}{3}\right)^{1/7} \mtot \frac{\eta^{1/7}}{w^{9/7}}.
\end{eqnarray}
Substituting in Eq.~(\ref{eq:rmin}), this corresponds to a maximum pericenter distance
\begin{eqnarray}
\label{eq:rpmax}
r_{p,\max} &=& \left( \frac{85 \pi}{6\sqrt{2}}\right)^{2/7} \mtot
  \frac{\eta^{2/7}}{w^{4/7}}\\\nonumber
&&\quad\times
\left(1-\frac{1}{4}\left( \frac{85 \pi}{3}\right)^{2/7}(4\eta)^{2/7}w^{10/7}\right).
\end{eqnarray}
We shall demonstrate below that event rates are dominated by the central regions of the
galactic cusp.
For $m = M$ ($\eta =1/4$) and $w= 1000\,\kms$, $b_{\max} \approx 2900\,\mtot$, and $r_{p{\max}}= 47\,\mtot$.
The correction in Eq.~(\ref{eq:rpmax}) is clearly negligible for nonrelativistic initial conditions.

\subsubsection{Eccentric Inspiral}
After the binary forms with dimensionless pericenter distance $\rho_{p0}=r_{p0}/\mtot$ and eccentricity $e_0$, its orbit will decay through the emission of
GWs \citep{\peters},
\begin{equation}
\frac{\rmd \rho_p}{\rmd t} =-\frac{64}{5} \frac{\mtot^{-1} \eta^2}{\rho_p^{3}}\frac{(1-e)^{1/2}}{(1+e)^{7/2}}\left(1+\frac{73}{24} e^{2}+\frac{37}{96} e^{4}\right)
\label{pet1}
\end{equation}
\begin{equation}
\frac{de}{dt} = -\frac{304}{15}  \frac{\mtot^{-1} \eta}{\rho_p^{4}}e\frac{(1-e)^{3/2}}{(1+e)^{5/2}}\left(1+\frac{121}{304}e^{2}\right)
\label{pet2}
\end{equation}
where we assume $\rho_{p}=r_p/\mtot= a/[\mtot(1-e)]$ until the last stable orbit.

We evolve the orbits following \citet{\peters} for the
evolution of the dimensionless periapsis $\rho_p$, the
time-to-merger, $t$, as a function of the instantaneous eccentricity,
$e$. Note that if starting from a parabolic passage, the initial condition
for the orbit is determined from exactly one parameter, $\rho_{p0}$.
Dividing Eq.~(\ref{pet1}) with Eq.~(\ref{pet2}) yields a separable differential
equation for $\rho_p(e)$. The solution satisfying a parabolic encounter initial
condition $\rho_{p}=\rho_{p0}$ at $e=1$ is
\begin{equation}\label{e:rho_p1}
\rho_p(\rho_{p0},e) = \rho_{p0} \kappa_{\rho}(e)
\end{equation}
where
\begin{equation}\label{e:rho_p2}
\kappa_{\rho}(e) = 2 \left(\frac{304}{425}\right)^{\frac{870}{2299}} e^{\frac{12}{19}}\left(1+\frac{121}{304}e^2\right)^{\frac{870}{2299}}
(1+e)^{-1}
\end{equation}
for which $\kappa_{\rho}=1$ at $e=1$.  The orbital evolution
equations~(\ref{e:rho_p1}) \& (\ref{e:rho_p2}) are valid far from the
horizon $\rho_p\gg 2$. Once the last stable orbit (LSO) is reached the
evolution is no longer quasiperiodic and the binary quickly coalesces. In the
leading order ratio approximation for {\new infinite mass ratio and} zero
spins,
\begin{equation}
  \label{eq:LSO}
 \rho_p(e_{\rm LSO}) = \frac{6+2e_{\rm LSO}}{1+e_{\rm LSO}},
\end{equation}
which we solve numerically for $e_{\rm LSO}(\rho_{p0})$ using
Eqs.~(\ref{e:rho_p1}) \& (\ref{e:rho_p2}).

The time evolution of the eccentricity can be written using Eqs.~(\ref{pet1})~\&~(\ref{pet2}) as
\begin{eqnarray}
 t_{\rm mg}(e) &=& \frac{15}{19} \left(\frac{304}{425}\right)^{\frac{3480}{2299}} \mtot \eta^{-1} \rho_{p0}^4 \nonumber\\
 &&\quad\times \int_0^{e} {\epsilon}^{\frac{29}{19}} \frac{\left(1+\frac{121}{304}{\epsilon}^2\right)^{\frac{1181}{2299}}}{\left(1-{\epsilon}^2\right)^{3/2}} \D \epsilon
\label{e:t-e}
\end{eqnarray}
where $t_{\rm mg}(e)$ denotes the time remaining until coalescence when the eccentricity is $e$.
Close to coalescence this is approximately,
\begin{equation}
 t_{\rm mg}(e) \approx \frac{5}{16} \left(\frac{304}{425}\right)^{\frac{3480}{2299}} \mtot \eta^{-1} \rho_{p0}^4 e^{\frac{48}{19}} {~~\rm if}~ e\ll 1.
\end{equation}
while the total merger time starting from a highly eccentric initial orbit \citep{\peters}
\begin{equation}
t_{\rm mg}\approx \frac{3}{85} \frac{a^4}{\mtot^3 \eta} (1-e^2)^{7/2} \text{~~if~} e\approx 1.
\label{timerge}
\end{equation}

Substituting Eqs. (\ref{eq:a}) \& (\ref{eq:e}) for the initial separation and eccentricity $a_0$ and $e_0$
into Eq. (\ref{timerge}) we get,
\begin{equation}
  \label{eq:fimerge}
  t_{\rm mg} \approx \frac{3 \sqrt{3}}{170 \sqrt {85 \pi}}
  \frac{(b w )^{21/2} }{\mtot^{19/2} \eta^{3/2} }
\end{equation}
where we have set $E_{\rm final} = \delta E$, ignoring the initial energy
of the binary.  This is an excellent approximation all the way out to $0.99
b_{\rm bmax}$ because of the strong dependence of $t_{\rm merge}$ on the
initial velocity $w$.
The merger time reaches its maximum at $b = b_{\max}$ and simplifies to
$4\pi G \mtot / w^3$: about two times the orbital period of a binary with
circular velocity $w$.

For the most relevant systems, we find that such binaries will merge before
they have a close encounter with a third body.  The binary can be disrupted
if it encounters a third body with a closest approach $\lesssim 2a$ before
it merges. For the maximum impact parameter to form a binary, $b_{\rm
max}$, the separation of the binary that forms is $a \approx G \mtot /w^2$.
The typical timescale for an encounter to disrupt the binary is $\sim (12
\pi a^2 n w)^{-1} = w^3/(12 \pi G^2 \mtot^2 n)$, where the additional
factor of $3$ comes from gravitational focusing.  Because of the strong
dependence Eq.~(\ref{eq:fimerge}) has on the impact parameter ($\propto
b^{21/2}$), for any reasonable disruption timescale, a slightly closer
impact will always compensate and allow the binary to merge before being
disrupted.  In fact, in order for the disruption timescale to be comparable
to merger timescale, the number density of stars must be $n \gtrsim
10^{15}\,$pc$^{-3}$ for $w = 100\,\kms$. Because every binary results in a
rapid merger, the rate of inspiral at any time is well approximated by the
rate of binary formation.

\begin{figure}
\centering
 \mbox{\includegraphics{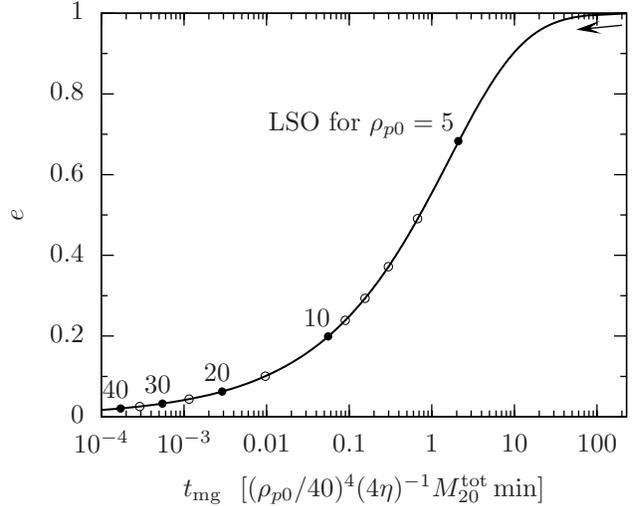}}
\caption{\label{f:t-e} The eccentricity evolution as a function of time to
merger. All of the orbits start off near $e=1$, and depending on the
initial periapsis $\rho_{p0}$, LSO corresponds to the eccentricity $e_{\rm
LSO}$ shown with filled circles as marked. Open circles give $e_{\rm LSO}$
for equidistant intermediate values for $\rho_{p0}$. The curves should not
be extrapolated to $e< e_{\rm LSO}$ as the evolution might be very
different during the BH merger.}
\end{figure}

Equation~(\ref{e:t-e}) provides an explicit monotonic relation between
time and eccentricity. This relation is sensitive to the free
parameters $\mtot$, $\eta$, and $\rho_{p0}$ only through the
overall scaling $\mtot \eta^{-1} \rho_{p0}^4$. Therefore
eccentricity can be thought of as a dimensionless time
variable. Figure~\ref{f:t-e} illustrates the conversion between $t$
and $e$. For the parameters, $\rho_{p0}=40$, $\mtot=20\Msun$, and
$\eta=0.25$, the characteristic timescale is several hours for
moderate--large eccentricities, and minutes for the final small
eccentricity inspiral.

\subsection{Event Rates}
\label{sec:rates}
Given the cross-section for binary formation ($\sigma_{\rm cs}=\pi
b_{\max}^2$), we can now calculate the expected rate of binary
formation in a single galactic nucleus, $\Gamma_{1\rm GN}$. In order
to evaluate the total contribution of different BH masses, $(m,M)$,
velocities, $(\bm{v}_m,\bm{v}_M)$, and spatial position, $r$, within a
single galactic nucleus, we need to integrate the differential rate
$\D^9 \Gamma_{1\rm GN}=\sigma_{cs} w f_m(r,\bm{v}_m)f_M(r,\bm{v}_M)$
over the corresponding distributions.  Here $w = | \bm{v}_m - \bm{v}_M
|$ is still the magnitude of relative velocity between the BHs, and
$f_m$ and $f_M$ are the 6-dimensional distribution functions of the
BHs with mass $m$ and $M$ derived in \S~\ref{sec:massseg}.
Generally, this calculation requires us to evaluate an integral over 9
variables,
\begin{eqnarray}
  \label{eq:integral}
  \Gamma_{1\rm GN} &=& \int_{r_{\min}}^{r_{\max}}\rmd r 4\pi r^2 \int_{M_{\min}}^{M_{\max}} \rmd M
  \int_{M_{\min}}^{M} \rmd m \\\nonumber
&& \quad     \int\!\!\!\int_{x_m, x_M > 10, J >J_{\rm LC}} \hspace{-1.8cm}\rmd^3 v_m \rmd^3 v_M f_m(r,\bm{v}_m)f_M(r,\bm{v}_M) \sigma_{\rm cs} w,
\end{eqnarray}
where the integration bounds are set consistent with
the distribution domains (see \S \ref{sec:massseg}).  The multi-dimensional
integration can be greatly simplified to only three variables using a
few approximations. First in \S 2.3, we assumed that the the velocity distributions
were isotropic. For fixed $m$ and $M$, we switch integration variables from
$(\bm{v}_m, \bm{v}_M)$ to $(\bm{v}_m-\bm{v}_M, \bm{v}_m+\bm{v}_M)$, and adopt
spherical coordinates. Since the integrand only depends on $w$, we can evaluate the
integrals over the other 5 velocity components\footnote{Some difficulties
arise because of the integration bounds depend on the other parameters. However,
the integrals can be evaluated under the approximation $r_{\min}\leq r \leq r_{\max}$ and $0\leq v\leq v_{\rm esc}(r)$,
where $r_{\min}$ and $r_{\max}$ are the minimum and maximum radii for a relaxed population of BHs in the nucleus
and $v_{\rm esc}(r)$ is the escape velocity at radius $r$.}
\begin{eqnarray}
\int \rmd^3 v_m \int\rmd^3 v_M\, f_m(r,\bm{v}_m)f_M(r,\bm{v}_M) \,\sigma_{\rm cs} w= \nonumber \\
=n_m(r) n_M(r) \int \rmd w \,  \psi_{mM}(r,w)\, \sigma_{\rm cs} w,
\label{eq:relvel}
\end{eqnarray}
where $\psi_{mM}(r,w)$ is the relative velocity distribution which can be
expressed as a one-dimensional integral using the Dirac-$\delta$ function for a power-law
dimensionless energy distribution profile $g_m(x)\propto x^{p_m}$ (see \S~\ref{sec:massseg}
for a definition). {\new Note} that the phase space distribution
functions of the BHs and stars are well approximated by a power-law
$g_m \propto x^{p_m}$ for $x > 10$.

We have numerically integrated the remaining velocity integral
over the possible range of $w$ for a variety of
different slopes $p_m$ and $p_M$ and find that to within $\lesssim
10\%$ the integrand is independent of the shape of the {\new relative} velocity distribution and
only depends on the expected value of the relative velocities, and can
be expressed with the expected value of the individual velocity magnitudes as
\begin{equation}
  \label{eq:simpint}
\int \rmd w \, \psi_{mM}(r,w) \,\sigma_{\rm cs} w
  \approx \pi b_{\max}^2 v_{\rm c}(r),
\end{equation}
where $b_{\max}^2$ is evaluated at $w = v_{\rm c}(r)$ using Eq.~(\ref{eq:bmax}), and $n_m(r) \propto r^{-p_m-3/2}$.
This is to be expected, since the encounters are practically parabolic such that the initial
velocity profile is negligible compared to the velocity at periastron.
Thus, Eq.  (\ref{eq:integral}) simplifies to
\begin{equation}
  \label{eq:diffsimpfin}
  \frac{\rmd^3 \Gamma_{1\rm GN}}{\rmd r \,\rmd m \,\rmd M}= 4 \pi^2 b_{\max}^2 v_{\rm c}(r) n_m(r)
  n_M(r) r^2,
\end{equation}
so that the total rate in one galactic nucleus is
\begin{equation}
  \label{eq:intsimpfin}
  \Gamma_{1\rm GN} = \int_{r_{\min}}^{r_{\max}} \!\!\rmd r \int_{M_{\min}}^{M_{\max}} \!\!\rmd M \int_{M_{\min}}^{M} \!\!\rmd m
\;\frac{\rmd^3 \Gamma_{1\rm GN}}{\rmd r \,\rmd m \,\rmd M}.
\end{equation}
In practice, we calculate $\Gamma$ for all of our simulations by calculating
$n_m(r)$ and $n_M(r)$ for discreet masses and sum over $m < M$.

\subsection{Results and Discussion}
\label{sec:properties}

The resulting event rates integrated over a single galactic nucleus
are presented in the sixth column of Table~\ref{table1} for all of our
model nuclei. The estimated rates vary between
$\sim 10^{-8}$--$10^{-10}\yr^{-1}$ over the various models.
We discuss the primary sources of uncertainties and other
important aspects related to the event rates in detail below.

\subsubsection{Number of Black Holes}
Overall, we find that the merger rate is most sensitive
to $C_{\rm BH} = n_{\rm BH}/ n_*$, however, the {\new accretion of BHs by
the SMBH} reduces the na\"{i}ve scaling relation $\propto
C_{\rm BH}^2$.  Thus, as the number fraction of BHs increases by a
factor of $100$, the rate of mergers only increases by a factor of
$20$.  The merger rate depends far less so on $M_{\max}$ and
$M_{\rm min}$, since, for larger $M_{\max}$ the total number of BHs
tends to be reduced in the inner $\sim 0.1\,$pc of the SMBH.  The
rate, however, remains relatively unchanged since the cross-section for
binary capture increases with the BH mass.

\subsubsection{Pericenter Distance Dependence}
\begin{figure*}
\begin{center}
  \includegraphics[width=\columnwidth]{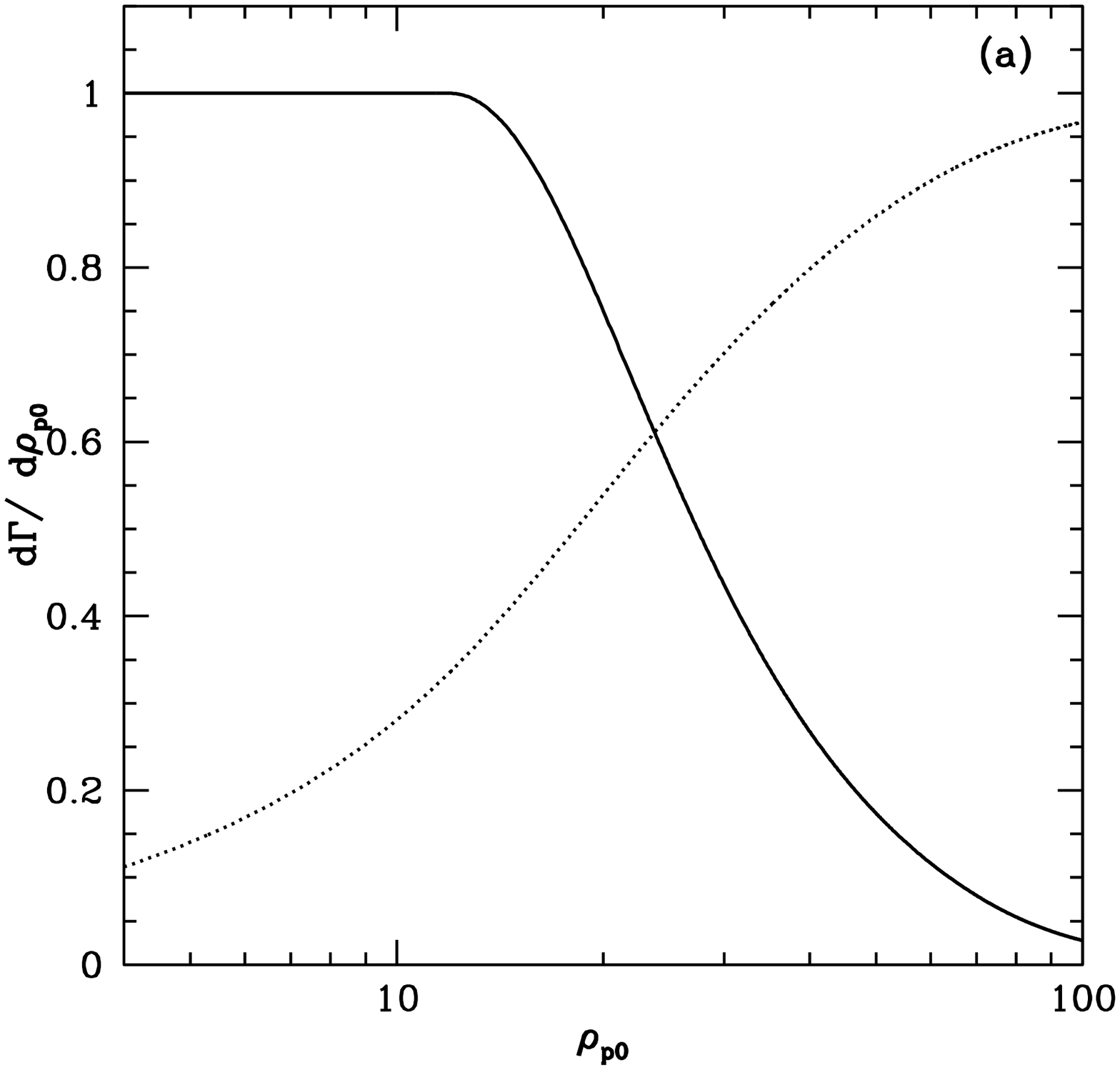}
  \includegraphics[width=\columnwidth]{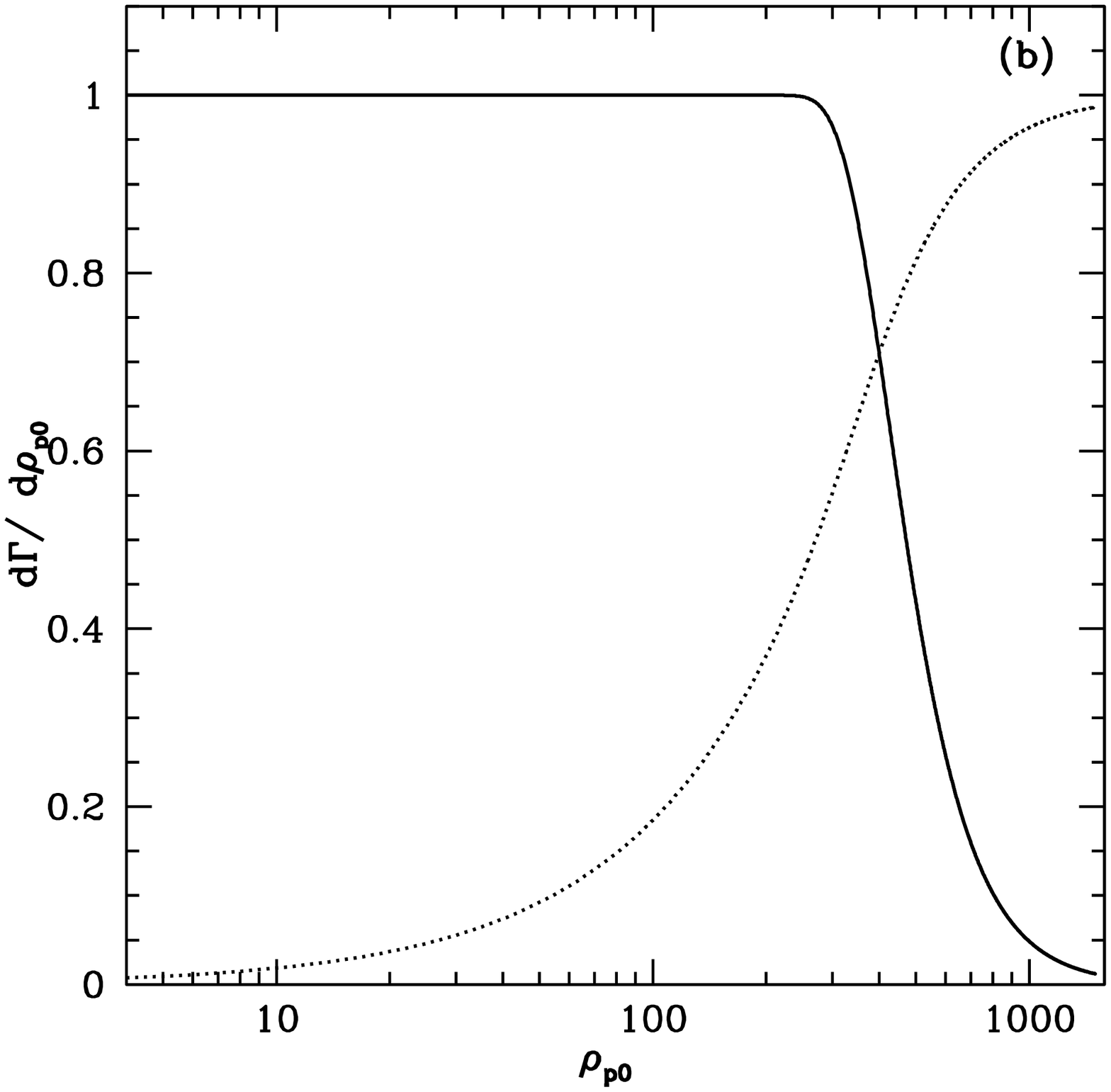}
\end{center}
\caption{\label{fig:distribution} Normalised probability distribution
  and integrated probability distribution of initial pericenter
  distances for binaries in galactic nuclei (a) and {\new in
    Spitzer-unstable} massive star clusters {\new without a central
    massive black hole} (b{\new ; see \S~\ref{sec:clusters} for
    details}).  To calculate the values in this figure, we have
  assumed that the merger rate is dominated by the $10\,\Msun$ BHs
  with a near Maxwellian velocity dispersion, and in galactic nuclei
  they have density profiles $\propto r^{-2}$, consistent with the
  density profile of the most massive objects in the nucleus (see
  \S~\ref{sec:massseg:results}).  The solid line is the differential
  rate of binary formation $\rmd \Gamma / \rmd r_p$ and the dotted
  line represents the integrated probability $\Gamma (r>r_p)$. For
  galactic nuclei, the probability distribution is perfectly uniform
  out to pericenter distance $r_{p\max}(w_{\max}(.001\,$pc$) \approx
  12 \mtot$, after which it drops off quickly.  For globular clusters
  {\new without a central massive BH} the distribution is uniform out
  to $r_{p}\approx 250 \mtot$, and it begins to drop off as a Gaussian
  profile for $r_p \ga 540 \mtot$.  }
\end{figure*}

For parabolic encounters, the differential of pericenter distances that
lead to binary formation, according to Eq.~(\ref{eq:rmin}) is to leading order
\begin{equation}
\rmd r_p \approx \frac{ w^2 b \rmd b}{\mtot},
\end{equation}
where we have fixed $w$, the relative velocity at infinity. Since
$\rmd \Gamma / \rmd b \propto b \propto \rmd r_p/ \rmd b $, the
pericenter distribution {\new that lead to binary capture ($\rmd \Gamma / \rmd
  r_p$) } is uniform out to a maximum pericenter distance $r_{p{\max}}$.  We can calculate the overall distribution of encounters
that form binaries by changing the order of integration of
Eq.~(\ref{eq:integral}) and leaving it as a function of $r_p$,
\begin{eqnarray}
\frac{\rmd \Gamma_{1\rm GN}}{\rmd r_p} &=& \int  \!\!\!  \int  \!\!\!  \int
\!\!\!  \int \rmd r \rmd  w\rmd m \rmd M\; 4 \pi r^2 n_M(r) n_m(r) \nonumber\\&&\quad\times\frac{\mtot}{w^2}  w \psi_{mM}(r,w),
\label{eq:dgdrp}
\end{eqnarray}
where $\psi_{mM}(r,w)$ is the distribution function of relative velocities given by Eq.~(\ref{eq:relvel}).
The limits of the integration determine the functional form in
three main regimes.  For $r_p < r_{p{\max}}[w_{\max}(r_{\min})]$ where
$w_{\max}(r)= 2 v_{\rm esc}(r) = 2\sqrt{2} v_c(r)$
is the maximum relative velocity at radius $r$, $v_{\rm esc}$ is the escape velocity, and $r_{p{\max}}(w)$ is defined in Eq.~(\ref{eq:rpmax}),
Eq.~(\ref{eq:dgdrp}) is independent of $r_p$ and is integrated over
the bounds
\begin{align}
0 < w {} < w_{\max}(r)  \nonumber \\
r_{\rm min} < r {} < r_{\max}.
\end{align}
For $r_{p\max}(w_{\max}(r_{\rm min})) <r_p< r_{p\max}(w_{\max}(r_{\max}))$
the limits of integration are determined by $w_{\rm
eq}(r_p)$, the inverse of Eq.~(\ref{eq:rpmax}) and $r_{\rm eq}(r_p) = 2 G
\msmbh/ w_{\rm eq}(r_p)^2$. In this regime, the limits of integration are
split into two regions
\begin{align}
0 < w  < w_{\rm eq}(r_p) \nonumber \\
r_{\rm min} < r {} < r_{\rm eq}(r_p),
\end{align}
and
\begin{align}
w_{\rm eq}(r_p) < w {} < w_{\max}(r) \nonumber \\
r_{\rm eq}(r_p) < r {} < r_{\max}.
\end{align}
Finally, for $r_p > r_{p{\max}} (2 \sqrt{2} v_c(r_{\max}))$ the
limits of integration are
\begin{align}
0 < w {} < w_{\rm eq}(r_p) \nonumber \\
r_{\rm min} < r {} < r_{\max}.
\end{align}
A normalised probability distribution of pericenter distances for
encounters in galactic nuclei is plotted in
Figure~\ref{fig:distribution}a, where we approximated $\psi_{mM}(r,w)$ as a
Maxwellian distribution with variance $v_c(r)$.
For comparison the distribution function for the Maxwellian core of a
star cluster {\new without a central massive BH} is plotted in Figure~\ref{fig:distribution}b.

\subsubsection{Eccentricity dependence}

\begin{figure}
\begin{center}
\includegraphics[width=\columnwidth]{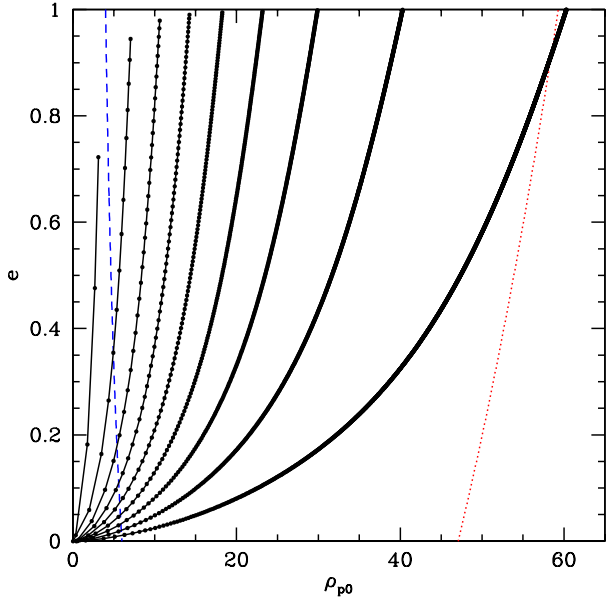}
\end{center}
\caption{\label{fig:inspirala} The secular evolution of BH binaries.
  We have plotted in $10\%$ probability intervals the evolution of the
  binaries as they decay ($M = m = 10\,\Msun$; see
  Fig.~\ref{fig:distribution} for details).  The solid line is the
  orbit averaged evolution given by
  Eqs.~(\ref{e:rho_p1})~\&~(\ref{e:rho_p2}).  The solid circles denote
  the completion of approximately one orbital period, $\Delta t_{\rm
    orb}$. The gap between the solid line and $e=1$ is due to the
  finite loss of energy during the initial parabolic encounter.  For
  nearly $\sim 30\%$ of all binaries, the orbit averaged approximation
  is (visibly) not valid, as can be seen by the large space between
  each orbit.  The dotted (red) line denotes where the binaries peak harmonic
  is $10\,$Hz, the lower limit of the LIGO band. Nearly $90\%$ of all
  binaries that form in galactic nuclei are within this limit upon
  first passage. The dashed (blue) line denotes the eccentricity at the last
  stable orbit (Eq.~\ref{eq:LSO}).  }
\end{figure}

We can use Eqs.~(\ref{e:rho_p1}), (\ref{e:rho_p2}), and (\ref{e:t-e})
to follow the secular evolution of the binary as it passes through the
LIGO band and merges when it reaches its last stable orbit.  In
Figure~\ref{fig:inspirala}, we have plotted the secular evolution of
formed binaries in $10\%$ probability intervals, and plotted the
eccentricity after approximately every complete orbit with small
circles. Nearly $90\%$ of all binaries actually form within the LIGO
band, and $\approx 94\%$ have an eccentricity $e>0.3$ as it enters the
LIGO band (when $f_p=10\Hz$).  This signature is unique to an active
cluster of BHs in a high velocity dispersion environment.  For lower
velocity dispersions as in star clusters, the {\new merger}
distribution of pericenter distances {\new ($\rmd \Gamma/ \rmd r_p$)}
stays uniform out to $\approx r_{p{\max}}(2\times 50\,\kms) \approx
260\,\mtot$. In contrast to galactic nuclei, we expect only $\approx
10\%$ of all binary GW sources {\new in star clusters} to have
eccentricities higher than $0.3$ when they enter the LIGO band, and
$\approx 8\%$ to form within the LIGO band.  Binaries which merge due
to 3 or 4--body interactions, are even less likely to have such a high
eccentricity \citep{2006ApJ...640..156G}.  \citet{2006ApJ...637..937O}
calculated the expected distribution of eccentricities for binaries
that form in dense star clusters (see their Fig.~3).  In their
simulations which had over $1000$ mergers from random encounters, they had no cases that had an eccentricity $>0.3$
{\new when the binary's GWs entered} the LIGO band, and only 3 cases
($\approx 10^{-3}$) with eccentricity $>0.1$. Their calculations also
included secular effects which could result in mergers with higher
eccentricity \citep{2003ApJ...598..419W}. However, in these cases, the
binary must have merged after one Kozai cycle in a hierarchical
triple. {\new Since the Kozai cycle is a dynamical effect, it operates
  on a timescale much shorter than the disruption timescale of the
  cluster.  Thus, any merger due to the Kozai effect would indicate an
  active cluster of BHs must still exist.  This is in contrast to a
  delayed merger from an ejected binary, which takes of order a Hubble
  time to merge. } We discuss other aspects of GW detection in
\S~\ref{sec:detection} below.

\begin{figure}
\begin{center}
\includegraphics[width=\columnwidth]{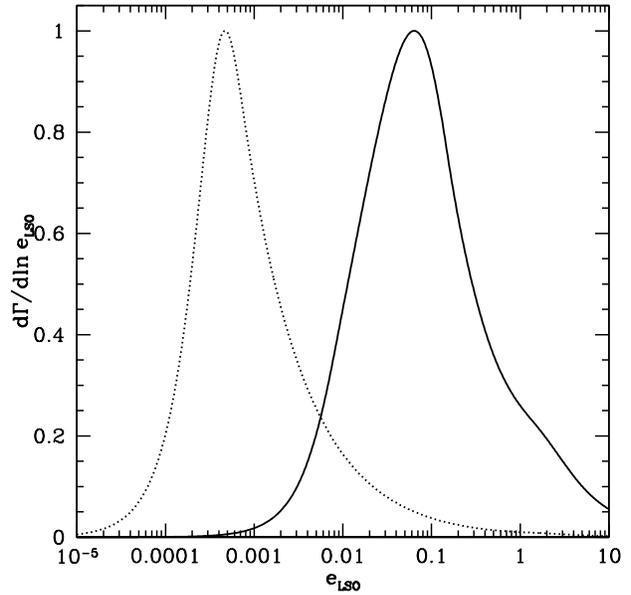}
\end{center}
\caption{\label{fig:dpdlne} The eccentricity distribution of events.
  Plotted is the eccentricity distribution of mergers at the last
  stable orbit ($\rmd \Gamma / \rmd \ln{e_{\rm LSO}}$) for one example
    of a galactic nucleus (solid line) and a globular cluster (dotted
    line).  Both lines are normalised so that they reach a maximum
    value of $1$.  $e_{\rm LSO} \ge1 $ corresponds to encounters that
    directly undergo a plunge. See the text and
    Fig.~\ref{fig:distribution} for our assumptions and details of the
    calculation.}
\end{figure}

In most instances, the first binary forming encounters occur
within the LIGO band. However, the S/N of such encounters is too small
to be detected for low masses \citep{2006ApJ...648..411K}. Therefore,
encounters which remain eccentric throughout the inspiral, especially
near plunge, may be the most readily detectable encounters.  Therefore
we are interested in the eccentricity of the binary as it reaches the
LSO (see Eq.~\ref{eq:LSO}). From Eq.~(\ref{eq:dgdrp}), and the
equation of evolution \citep[][ Eq.~\ref{e:t-e}]{1964PhRv..136.1224P}
we solve for the probability distribution of eccentricity at the LSO.
We plot the eccentricity distribution at LSO in Figure~\ref{fig:dpdlne}
for both galactic nuclei and globular clusters.
For encounters which that
to direct plunge, our calculation gives an eccentricity greater than 1.
However, for normalisation purposes, we include this in our
calculations, as they interestingly comprise a significant fraction of
merger events.

Until now, nearly all LIGO sources were expected to have a negligible
eccentricity as they enter the LIGO band \citep[but see][for intermediate mass ratio inspirals in star clusters]{2008ApJ...681.1431M}.
The comparatively low eccentricity binary formed through few-body
encounters or standard binary evolution
circularise before they enter the LIGO band and are detected.
Therefore, the detection of eccentric inspirals is a strong test of the
formation scenario of nuclear binaries, and can conclusively reveal the
origin of the BHs.

\subsubsection{Radius Dependence inside the Galactic Nucleus}

\begin{figure*}
  \centering
  \includegraphics[width=\columnwidth]{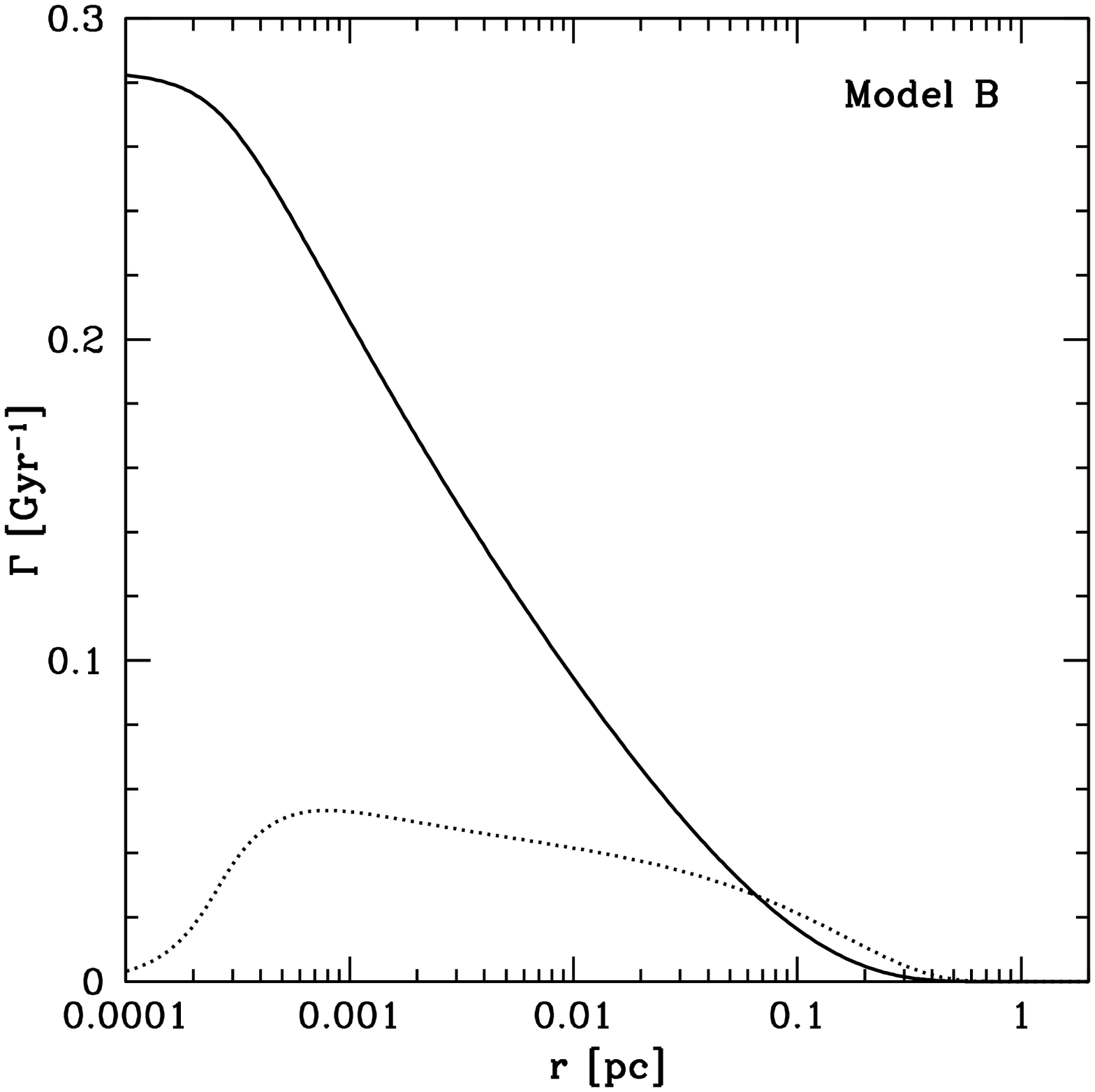}
  \includegraphics[width=\columnwidth]{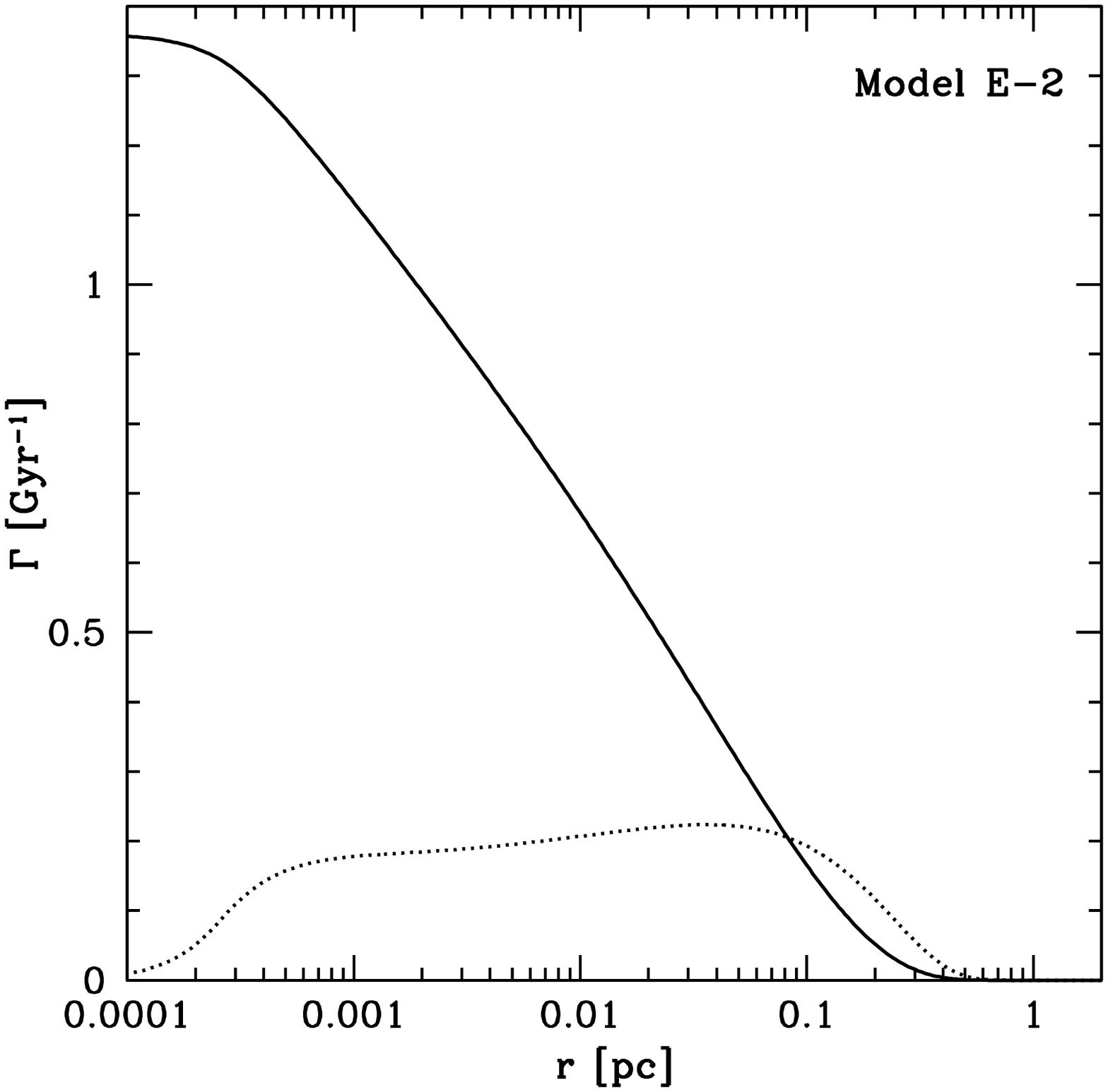}
  \caption{Rate of binary formation as a function of the log of the
    radius for Models B and E-2.  The solid line is the cumulative rate
    of binary formation for radii greater than $r$.  The dotted line
    is the differential rate distribution per logarithmic bin ($\rmd
    \Gamma / \rmd\log{r}$).}
  \label{fig:binformrate}
\end{figure*}

In Figure~\ref{fig:binformrate}, we plot the cumulative binary formation
rate for radii larger than $r$, $\Gamma(>r)$, as well as $\rmd \Gamma /
\rmd \ln{r}$. For most models, the total differential rate of binary formation
per logarithmic bin is roughly flat.
Thus, each logarithmic radius interval contributes equally to the rate.
We therefore conclude that the rates
determined are rather robust to the depletion of BHs very close to the
SMBH {\new as may be caused by resonant relaxation  \citep{1996NewA....1..149R, 1998MNRAS.299.1231R,2006ApJ...645.1152H}}
or our choice of the innermost radius for BHs. In order for the rate to be
dominated by mergers at large $r$, the number density of the BHs would have
to decrease with an exponent $r^{-\alpha}$ where $\alpha = p + 3/2 <3/2$
(see \S~\ref{sec:massseg}).
This is precisely the reason we accounted for the stars in determining
the potential in
Eq. (\ref{eq:numdens}),
and did not let the density profile of the BHs and
stars go to a constant value as in previous analyses.

We do not expect these tight binaries in their subsequent inspiral phase to
have any observable effect dynamically.  Overall, we expect $\sim 10 -
10^3$ such binaries to merge over a Hubble time.  This presents a much
smaller source of energy than the SMBH, which accretes $\sim 10^4$ BHs
over a Hubble time (FAK06).  However small the intrinsic rate in each
galaxy, the cumulative merger rate of many galaxies is large enough to be
detected by future ground based gravitational wave observatories.

\subsubsection{Cosmological Merger Rate Density}
\label{sec:cosmergerrate}

Using the $\msmbh-\sigma_*$ relationship found by \citet{2002ApJ...574..740T}
for higher mass BHs,
\begin{equation}
  \label{eq:msigma}
  \msmbh \approx 1.3\times 10^8\,\Msun ( \sigma_* /
  200\,\kms)^4,
\end{equation}
we can extrapolate our results from \S~\ref{sec:binform} to a range of
galactic nuclei and determine the overall rate of mergers in the
Universe.  {\new Observations by \citet{2005ApJ...619L.151B} \&
  \citet{2006ApJ...641L..21G} have demonstrated that
  Eq.~(\ref{eq:msigma}) extends to active SMBHs with masses as low as
  $10^5\,\msun$, well within the range of our interest.  }  We expect
the total mass of stars within the radius of influence to be
comparable to the mass of the SMBH, $M(< r_i) \approx 2 \msmbh$.  If
these stars follow a radial density profile of $n_* \propto r^{-3/2}$
as expected for a relaxed system (see \S~\ref{sec:massseg}), then
their number density at the radius of influence is
\begin{equation}
  \label{eq:nstar}
 n_*(\msmbh) \approx
1.2\times10^5\,{\rm pc}^{-3} \sqrt{10^6\,\Msun/\msmbh}.
\end{equation}
This gives a number density about $45\%$ larger than assumed in
\S~\ref{sec:massseg} for the Milky Way {\new and used in our calculates presented in Table~\ref{table1}}, but is well within the
expected scatter in number densities as we discuss later in this
section. {\new The actual number density at the radius of influence
  for a BH will depend on both the formation history of the galaxy as
  well as the merger history of the SMBH. } Because the merger rate is
usually greatest at the smallest radii, we can can determine the rate
in any nucleus by scaling Eq.~(\ref{eq:intsimpfin}) and evaluating the
integral at $r_{\rm min} \propto \sqrt{\msmbh}$, the radius where the
merger timescale of the BH into the SMBH is approximately a Hubble
timescale (approximately the inner radius of BHs).  Coincidentally,
this radius has the same scaling relation to the outer radius, $r_{\rm
  max} = r_i \propto \sqrt{\msmbh}$ {\new where we use
  Eq.~(\ref{eq:msigma}).}  The total merger rate, $\Gamma$ is simply
proportional to $\propto n_{\rm BH}^2 (b^2 w) r^3$ evaluated at $r =
r_{\rm min}$.  We can approximate the number density of BHs as $n_{\rm
  BH}(r) \approx C_{\rm BH} n_*(r_i) (r/r_i)^{-p-3/2}$. Substituting
in $r_i$, we get $n_{\rm BH}{\new (r_i)} \propto \msmbh^{-1/2}$,
independent of the power-law distribution of BHs, $p$.  The
cross-section of binary capture times the relative velocity is $\pi
b^2 w \propto w^{-11/7} \propto \msmbh^{-11/28}$, given $w \propto
(\msmbh/r)^{1/2}$.  Combining these dependencies, we {\new find the
  merger rate has a} relatively weak dependence on the mass of the
SMBH, $\Gamma \propto \msmbh^{3/28}$.  Over two orders of magnitude in
mass, we expect the rate to change only by $\approx 40-60\%$.  To test
this {\new scaling} relationship we have run a simulation with $\msmbh = 10^5\,\Msun$
and $\sigma_* = 30\,\kms$, and found that the rate was in fact
comparable to the relationship found here.  Any discrepancy is likely
due to the slight difference in the capture rate of the BHs and stars
(see Eq.~[\ref{eq:losscone}]).

Although the merger rate of BHs is not very sensitive to the SMBH
mass, it is sensitive to the intrinsic scatter of $n_*$ {\new for each
  galaxy}, which is used to normalise the entire distribution of of
the BHs {\new (see Eq.~[\ref{eq:numdens}])}. The expected rate of mergers is determined by $\langle
n_*^2\rangle$, the average of the number densities squared at the
radius of influence {\new for the distribution of galaxies in the
  universe}.  Given $\langle n_*^2\rangle = \langle n_*\rangle^2 +
\sigma_n^2$, where $\sigma_n$ is the variance of the number density at
the radius of influence {\new of a population of SMBHs}, we must
rescale our results by a factor
\begin{equation}
  \label{eq:xi}
  \xi = \frac{\langle n_*^2\rangle}{\langle n_*\rangle^2} = 1+\frac{\sigma_n^2}{\langle n_*\rangle^2}.
\end{equation}
{\new Previous studies on GW event rate estimates neglected corrections
due to cosmic variance, Eq.~(\ref{eq:xi}). However it is quite plausible 
that {\it not} all galaxies with the same SMBH mass have exactly the same 
number of stars in the central cusp. In fact,}
\citet{2007ApJ...671...53M} determined the relaxation times for
galaxies in the ACS Virgo Survey \citep{2004ApJS..153..223C}, and
found a relatively tight correlation, however, there was still
significant scatter {\new above} the mean by about an order of
magnitude. We have taken the results from \citet{2007ApJ...671...53M}
(specifically all nuclei with $\sigma_* < 140\,\kms$ in their Fig.~1),
and determined that $ \sigma_n^2/\langle n_*\rangle^2 \sim 30$.  {\new
  Although there is considerable uncertainty in the actual value of
  $\xi$, both observationally and theoretically, we expect the merger
  rate to be larger than we have so far calculated by $\xi\gtrsim
  10$--$100$, and scaled our results by $\xi_{30} = \xi / 30$. In our
  calculations, however, we use the conservative slope of
  Eq.~(\ref{eq:nstar}) to determine the mean value of $n_*$, and not
  the shallower slope found by \citet{2007ApJ...671...53M}, which
  would give rate estimates to be an order of magnitude larger than
  found here.}

We calculate the average cosmological merger rate density by
convolving the rate per galaxy with the number density distribution of
SMBHs in the universe.  We extrapolate the results of
\citet{2002AJ....124.3035A}, who found the best fit
number density distribution of massive SMBHs to be
\begin{equation}
  \label{eq:smbhnumber}
 \frac{\rmd n_{\rm SMBH}}{\rmd \msmbh} = c_o\left(\frac{\msmbh}{M_\bullet}\right)^{-1.25} e^{-\msmbh /M_\bullet},
\end{equation}
assuming this formula is valid all the way to $\msmbh=10^{4}\,\Msun$,
where $c_o = 3.2\times 10^{-11}\,\Msun^{-1}\,{\rm Mpc}^{-3}$ and
$M_\bullet = 1.3\times 10^8\,\Msun$.  Finally, we get the cosmological
merger rate by integration,
\begin{eqnarray}
  \label{eq:cosrate}
  {\cal R} &=& \int\limits_{10^4\,\Msun}^{10^7\,\Msun} \Gamma_{1\rm GN}(\msmbh) \frac{\rmd n_{\rm SMBH}}{\rmd
    \msmbh}  \rmd \msmbh \nonumber\\&\approx& 3\, \Gamma_{1\rm GN} \xi_{30}\,{\rm Mpc}^{-3},
\end{eqnarray}
where $\Gamma_{1\rm GN}$ is the expected rate of mergers for a single
galactic nucleus of a specific model shown in Table~\ref{table1}.  The
normalisation $3\xi_{30}\,{\rm Mpc}^{-3}$ follows from the
distribution given in Eq.~(\ref{eq:smbhnumber}), and also accounts for
the intrinsic scatter of $n_{*}$ {\new for a population of galaxies},
Eq.~(\ref{eq:xi}).  For our fiducial model of the MW, Model B, we get
a comoving rate density of $8.4\times 10^{-10} \xi_{30}\,{\rm
  yr}^{-1}\,{\rm Mpc}^{-3}$.  Because this integral is nearly flat in
the log of $\msmbh$, it results in a similar rate density per
logarithmic mass bin and {\new is not very sensitive to the limits of
  the integration. Taking the lower limit of currently observed SMBH
  masses, of $10^5\,\msun$, we get a rate $\approx 60\%$ of what we
  calculated here.  However, recent observations by \citet{2008ApJ...688..159G}  have shown that the $\msmbh-\sigma$ relation extends to
  the smallest SMBHs observed, some of which reside in galaxies
  without a classical bulge. Hence, } there is still $\approx 50\%$
uncertainty in the merger rate due to the true function of
$\Gamma_{1\rm GN}(\msmbh)$ discussed above, but the total rate density
is relatively robust given the other uncertainties in our calculation,
especially $\xi_{\rm 30}$ and $C_{\rm BH}$.

For future calculations, we define $\rmd {\cal R}_{mM}/\rmd
{\rho_{p0}}$ as the rate density for fixed masses $m$ and $M$
analogous to Eq.~(\ref{eq:diffsimpfin}) and (\ref{eq:dgdrp}).

\subsubsection{Application to Massive Star Clusters}
\label{sec:clusters}

BHs in massive star clusters {\new without a central massive black
  hole} will also undergo an epoch of mass segregation
\citep{1993Natur.364..421K,1993Natur.364..423S}, in which the BHs
segregate to the cluster core, and effectively decouple from the stars
forming their own subcluster
\citep{2000ApJ...528L..17P,2004ApJ...608L..25M,2006ApJ...637..937O,2008arXiv0804.2783M}.
This BH subcluster will continue to interact only with the BHs until a
sufficient number of BHs are ejected dynamically, that they come back
into equilibrium with the stars. This occurs approximately when
$N_{\rm BH} \lesssim 100$ \citep{2000ApJ...539..331W}.  For massive
clusters, with $\langle w_{\rm BH}\rangle \approx 15\,\kms$ and
$n_{\rm BH} \sim 10^6\,$pc$^{-3}$, the cluster evaporates before about
$\lesssim 10^8-10^9\,$yr. Scaling Eq.~(\ref{eq:intsimpfin}) to these
parameters, the cluster will have a BH-BH merger rate of
\begin{eqnarray}
\Gamma_{1\rm MSC} &\approx& N_{\rm BH} \left< n_{\rm BH} w_{\rm BH} \sigma_{cs} \right> \nonumber\\&\approx&
1.8\times 10^{-8}\,{\rm yr}^{-1} \times\left(\frac{N_{\rm BH}}{1000}\right)
\left(\frac{n_{\rm BH}}{10^6\,{\rm pc}^{-3}}\right)\nonumber\\
&&\quad \times\left(\frac{w_{\rm BH}}{15\,\kms}\right)^{-11/7},
  \label{eq:clusterrate}
\end{eqnarray}
during this period of evolution, {\new if the number density of BHs is
uniform in radius}.
This is comparable to, but slightly less than merger the rate found by
\citet{2006ApJ...637..937O} for the early evolution of a cluster of
BHs due to three-body and four-body encounters alone.  The detection
rate of such early mergers depends on the number of {\em young}
clusters (with $t_{\rm age} \lesssim 10^8-10^9\,$yr) within the
detection limit of LIGO.  Globular clusters, an important source of
delayed mergers, are too old to still have a BH subcluster.  However,
the young clusters in star burst galaxies would be an excellent source
if they survive sufficiently long in their hosts to undergo this
process of mass segregation \citep{2007PhRvD..76f1504O}.

{\new The eccentricity distribution of binary capture mergers in star
  clusters is plotted in Figure~\ref{fig:dpdlne}.  Overall, the rate
  of mergers in star clusters is dominated by small eccentricity
  events, which would be detected as circular inspirals by
  ground-based gravitational wave detectors.  However,} young star
clusters may have as many, or even more, eccentric mergers than are
expected in the nuclei of galaxies {\new if there are a sufficient
  number of mergers in young star clusters}.  In massive star
clusters, we expect $\sim 10\%$ of all gravitational wave captures to
merge with eccentricities similar to those in galactic
nuclei. Therefore, the distribution of low eccentricity events will be
indicative of the source of BH-BH mergers, and may be useful in
constraining the distribution and evolution of BHs in both galactic
nuclei and massive star clusters.

\section{Detection of gravitational waves}
\label{sec:detection}

To determine the expected detection rate of sources, we must now
calculate the maximum luminosity distance to which these inspirals are
detectable. In this section, we discuss the general properties of the
waveform, calculate the maximum distance of detection, and add up the
total expected detection rate for second generation terrestrial GW
instruments.

\subsection{General Properties of the Inspiral}
The evolution of the binary and the GW signal can
be separated into three phases:

\noindent
{\bf [I]} Highly eccentric encounters -- train of distinct GW bursts
  in time, broadband signal in frequency.

\noindent
{\bf [II]} Moderate-small eccentricity inspiral -- continuous GW
  signal in time, dominated by distinct frequency harmonics.

\noindent
{\bf [III]}  Merger and ringdown -- short duration peak GW power and
  exponential decay.

\begin{figure}
\centering
 \mbox{\includegraphics{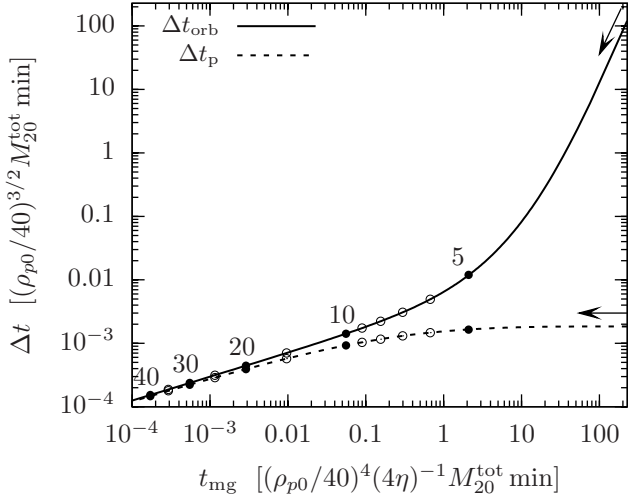}}
\caption{\label{f:t-dt} The relevant timescales that determine the GW
waveform: the evolution of the orbital time ($\Delta t_{\rm orb}$) and the
time duration of pericenter passage ($\Delta t_p$) as functions of time to
merger ($t_{\rm merger}$). The units are minutes on both axes for
$(M,\eta,\rho_{p0})=(20,.25,40)$ and are different for other values as
marked. Similar to Fig.~\ref{f:t-e} the orbits should not be extrapolated
beyond the LSO for a particular $\rho_{p0}$, shown with circles. The
inspiral is quasiperiodic if $\Delta t_{\rm orb}\ll t_{\rm merger}$ and
$\Delta t_p\ll t_{\rm merger}$. The GW signal is burstlike if
$\Delta t_p \ll \Delta t_{\rm orb}$ and is continuous if $\Delta t_p \sim
\Delta t_{\rm orb}$.}
\end{figure}

The distinction between Phases {\bf I} and {\bf II} can be understood
by studying the evolution of the relevant timescales that determine
the GW waveform: the orbital time ($\Delta t_{\rm orb}=\omega_{\rm
  orb}^{-1}$), and the time duration of pericenter passage ($\Delta
t_p=\omega_{p}^{-1}$).  These are plotted in Figure~\ref{f:t-dt} as a
function of time to merger $t_{\rm mg}$.  For both Phases~{\bf I} and
{\bf II}, $|\rmd \Delta t_{\rm orb}/\rmd t_{\rm mg}|\ll 1$, so the
orbit is quasiperiodic and evolves gradually due to the emission of
GWs.
The characteristic GW emission timescale during each orbit is determined by
$\Delta t_p$. Figure~\ref{f:t-dt} shows that initially (Phase~{\bf I}), $\Delta t_p\ll \Delta t_{\rm orb}$, implying
that the waveform consists of a train of short $\Delta t_p$ duration bursts arriving quasiperiodically
with separation $\Delta t_{\rm orb}$. Later, when the burst duration timescale $\Delta t_p$ becomes comparable to
$\Delta t_{\rm orb}$, the signal becomes continuous in time domain (Phase~{\bf II}).
Since the orbital evolution is quasiperiodic, the GW signal
is approximately a sum of discrete frequency harmonics of the orbital frequency, $f_{\rm
orb}$. When the eccentricity becomes relatively small $e \lesssim
0.7$, (Phase~{\bf II}) the harmonic decomposition is quickly convergent, while
during Phase~{\bf I}, it is more convenient to work with the continuous limit
of the frequency spectrum.

During Phase~{\bf I} and {\bf II} the equations of motion and the GW
 waveforms can be calculated {\new accurately in
 various approximations (see \S~\ref{sec:motivation} above)}.
{\new For Phase~{\bf II}, the expected signal to noise ratio of detection
has been calculated  by \citet{\bc}, and the $e=1$ parabolic case was
examined in \citet{2006ApJ...648..411K}. In the following, we
generalise these studies to be applicable to Phases~{\bf I}
and {\bf II}.}

 Once the orbital separation reaches the last stable orbit, the BHs
 fall in rapidly and form a common horizon. The last stable orbit,
 which marks the end of Phase~{\bf II}, is determined by the initial
 pericenter distance $\rho_{p0}$ for an initially parabolic orbit
 shown by circles in Figure~\ref{f:t-dt}.  The GW waveform during
 Phase~{\bf III} involves the calculation of the violently changing
 spacetime and the eventual relaxation into a Kerr BH. This requires
 full numerical simulations of the Einstein equations. The
 detectability of the resulting waveforms have been examined for
 nonspinning binaries and quasicircular initial conditions
 \citep{\baker,\berti}.  These studies have shown that $S/N=10$ can be
 reached up to a distance $d_L=1$ -- $6\,$Gpc for total binary mass
 $\mtot = 10$--$200\Msun$ for AdLIGO.  Eccentric mergers were
 considered very recently by \citet{\hinderb,\hindera} and
 \citet{\w} for the case of no initial spins, who showed that the
 resulting GW power is comparable to (or sometimes larger than) the
 power released during quasicircular mergers.
Future studies should address spin effects during the coalescence,
 they might significantly modify the GW
 power and waveforms.

 We note that the separation between Phases {\bf I}-{\bf II}-{\bf III}
 is valid only if the initial encounter has a minimum separation,
 $r_p$, that is much larger than the unstable circular orbit, $r_{\rm
   UCO} \sim 2$--$4\mtot$, depending on BH spins. Direct captures, or
 orbits outside but repeatedly approaching the unstable circular orbit
 (the so-called `zoom-whirl' orbits), are qualitatively
 different. Such encounters have been studied in the geodesic
 approximation appropriate for extreme mass ratios
 \citep{\gairof,\gairos,\pk,2008PhRvD..77j3005L} and using full numerical simulation for
 equal masses \citep{\hinderb}. In this case the GW spectrum is
 considerably different and the power is considerably increased.

The purpose of this section is to derive the signal-to-noise ratio for
the quasiperiodic Phases {\bf I} and {\bf II} and determine the maximum
range of detection for second generation terrestrial GW instruments.
We leave the assessment of Phase~{\bf III} and the zoom-whirl domain to future studies.

\subsection{Signal to Noise Ratio}

{\new Here we briefly review the general calculation of the signal to noise ratio for
detecting the GW signal, which we can then utilise for the waveforms generated by
GW capture events. We refer the reader to \citet{1998PhRvD..57.4535F} for more
details.}

In general, the signal to noise ratio of a GW detection is defined as
\begin{equation}\label{sn}
\frac{S^2}{N^2} = 4 \int_{f_{\min}}^{f_{\max}}\frac{|h^2(f,{\bm \theta})|}{S_h(f)} \D f,
\end{equation}
where $h(f, {\bm \theta})$ is the Fourier transform of the GW signal
weighted by the antenna beam patterns, ${\bm \theta}=\{\theta_{i}\}$ are
the physical parameters describing the source and the detector orientation,
$S_h(f)$ is the one-sided noise spectral density in units of $\Hz^{-1}$,
and $f_{\min}\leq f \leq f_{\max}$ correspond to the frequency band of the
instrument, e.g. $(f_{\min},f_{\max})\approx (10, 10^4)\,$Hz for AdLIGO\footnote{http://www.ligo.caltech.edu/advLIGO/scripts/summary.shtml}.

The sky position and binary orientation averaged root mean square
signal-to-noise for a single orthogonal arm interferometric GW
instrument is
\begin{equation}\label{sn2}
  \left\langle \frac{S^2}{N^2}\right\rangle = \int_{f_{\min}}^{f_{\max}}\frac{h^2_{c}(f)}{5 f S_h(f)}\, \frac{\D f}{f},
\end{equation}
where $h_{c}(f)$ is the characteristic isotropic GW amplitude
defined as
\begin{equation}\label{hc}
h_{c} = \frac{1}{\pi d_L}\sqrt{2 \frac{\D E}{\D f}} = \frac{1}{\pi d_L}\sqrt{\frac{2\dot E}{\dot f}}\, ,
\end{equation}
where $d_L(z)$ is the luminosity distance to a source at a cosmological
redshift $z$, $\D E/\D f$ is the one-sided GW energy spectral density on a
spherical shell at infinity. The second equality corresponds to the
stationary phase approximation for a quasiperiodic signal sharply peaked at
frequency $f$. In this case, $\dot E$ is the total GW power at frequency
$f$, which evolves slowly in time according to its time derivative $\dot f$.

Equations~(\ref{sn2}) and (\ref{hc}) can be generalised for a signal
consisting of discrete harmonics $f_n$, with negligible overlap as
\citep{\bc}
\begin{equation}\label{sn3}
\left\langle \frac{S^2}{N^2}\right\rangle =
\sum_{n=2}^{\infty}\int_{f_{\min}}^{f_{\max}}\frac{h^2_{c,n}(f_n)}{5 f_n S_h(f_n)}\, \frac{\D f_n}{f_n},
\end{equation}
and
\begin{equation}\label{hcn}
h_{c,n} = \frac{1}{\pi d_L}\sqrt{\frac{2\dot E_n}{\dot f_n}}\, ,
\end{equation}
where $\dot E_n$ is the GW power radiated at frequency $f_n$. For
quasiperiodic orbits with an intrinsic orbital frequency $f_{\rm orb}$, the
observed frequency harmonics at redshift $z$ are given by
\begin{equation}
\label{eq:forb}
f_n \equiv n f_{{\rm orb},z}\equiv n \frac{f_{\rm orb}}{1+z}.
\end{equation}

\subsection{Application to the GW Capture Process}
\label{sec:luminositydistance}

Let us now turn to the detectability of the GWs
starting from the initial hyperbolic encounter and ending
in the violent BH merger. {\new Here we
derive a computationally more efficient equivalent form of Eq.~(\ref{sn3}),
which can be utilised for Phases~{\bf I} and {\bf II}.}

\begin{figure*}
\centering
 \mbox{{\includegraphics{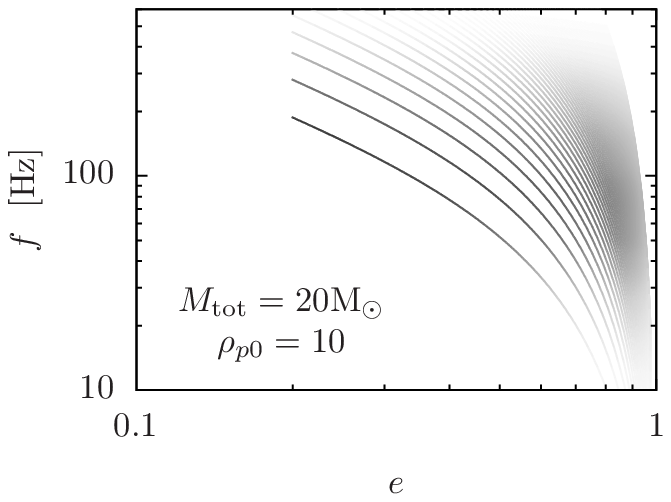}}
 {\includegraphics{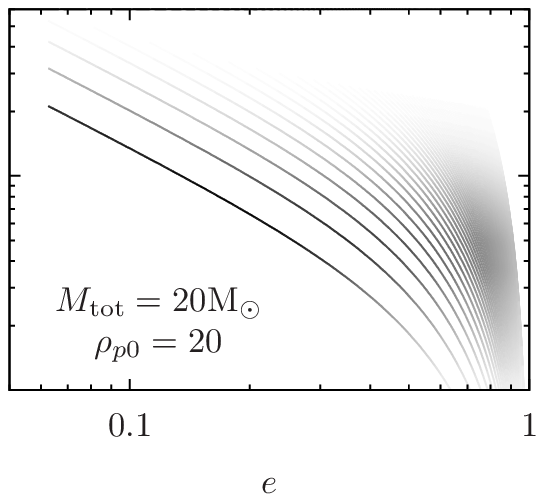}}
 {\includegraphics{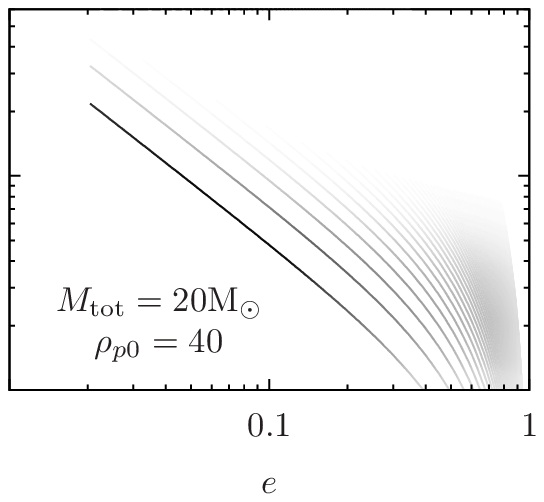}}}\\
\mbox{{\includegraphics{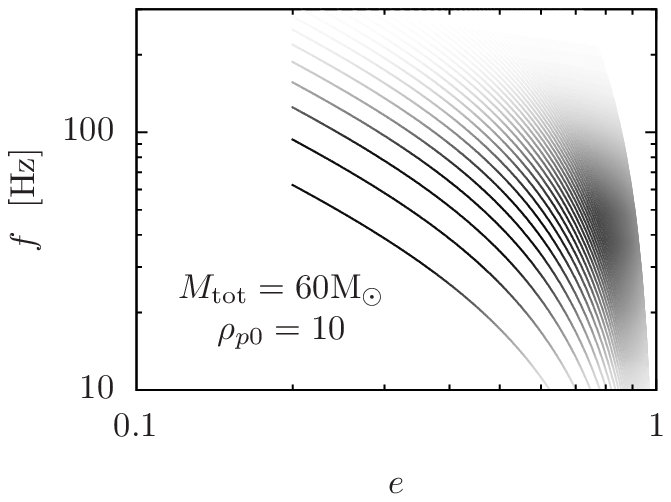}}
 {\includegraphics{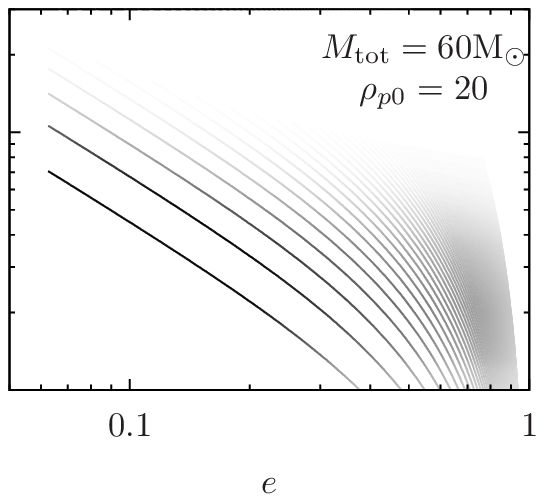}}
 {\includegraphics{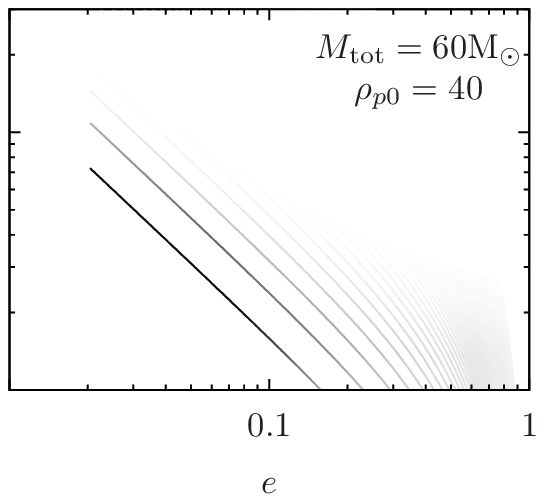}}}
\caption{\label{f:snr-eccentricity-frequency} The frequency evolution of
GWs as a function of eccentricity for various total mass and initial
periapsis as labelled.
The shading represents the expected signal-to-noise ratio of the first
$n=100$ harmonics per logarithmic eccentricity bins for AdLIGO. The
waveform is described by a broadband spectrum at large $e$ (Repeated Burst
Phase) that later separates into distinct harmonics as eccentricity
decreases until the LSO (Eccentric Inspiral Phase). The signal-to-noise
ratio is substantial already at $e\gtrsim 0.7$. }
\end{figure*}

{\new Following {\citet{\bc}}}, we calculate the binary evolution and the GW signal waveform in the
leading order approximation of \citet{\pmat}, where the interacting
masses move on quasi-Newtonian trajectories and emit quadrupolar
radiation.  This approximation is adequate in terms of the angular-averaged
$S/N$ if the initial periapse is well outside the unstable circular orbit, e.g.
$r_p \gg r_{\rm UCO}$. At smaller initial $r_p$, the Newtonian
approximation underestimates the GW power and is therefore
conservative.\footnote{{\new Modulations due to general relativistic pericenter
precession and relativistic beaming would be very important for the real data analysis, but
not in terms of the the calculation of event rates that depend on the angular-averaged total $S/N$.}}
To lowest order,
\begin{equation}\label{dotEn}
\dot E_n = \frac{32}{5} \eta^2 \mtot^{10/3}\omega_{\rm orb}^{10/3} g(n,e),
\end{equation}
where $g(n,e)$ determines the relative power of the $n$th harmonic for
orbits with eccentricity $e$, which is given by \citet{\pmat} in the
Newtonian approximation as
\begin{eqnarray}\label{gne}
g(n,e) &=& \frac{n^4}{32} \left\{\left[ J_{n-2} - 2eJ_{n-1} + \frac{2}{n}J_{n}
+ 2e J_{n+1} - J_{n+2} \right]^2 \right.\nonumber \\
&&\left.
 + (1-e^2)[ J_{n-2}- 2J_{n} + J_{n+2}]^2  + \frac{4}{3n^2}J_{n}^2
\right\}.
\end{eqnarray}
Here $J_i(x)$ is the $i$th Bessel function, and we have suppressed the argument
$x=n e$, i.e. $J_i \equiv J_i(ne)$ for each $i$ above. These waveforms generally
have a maximum at frequency $\omega_p$ for $e\gsim 0.5$ and have a steep cutoff
such that the fractional GW power beyond $5\omega_p$ is smaller than $10^{-3}$
\citep{\tu}.  For smaller $e$, the harmonics beyond the first $2\omega_p$ are
greatly suppressed. The number of harmonics necessary to a precision $10^{-3}$
can be estimated as
\begin{equation}\label{nmax}
 n_{\max} = 5 \frac{\omega_{p}}{\omega_{\rm orb}} = 5 \frac{(1+e)^{1/2}}{(1-e)^{3/2}}
\end{equation}
e.g. $n_{\max}=(10,40,10^4)$ for $e=(0.3,0.7,0.99)$, respectively.

Note that Eq.~(\ref{dotEn}) depends explicitly on $e$. Therefore in
order to directly evaluate the $S/N$ integral in Eq.~(\ref{sn3}) over
$f_n$ one needs to invert the frequency evolution equation
$f_n(r_{p0},e_0,e)$. Computationally it is much more efficient to
change the integration variable from $f_n$ to $e$,
\begin{equation}
\frac{\D f_n}{f_n} = - \frac{3}{2}\D\ln \frac{a}{a_0} =
-\frac{18}{19}\frac{1 + \frac{73}{24}  e^2 + \frac{37}{96}  e^4}{
                    1 - \frac{183}{304}e^2 - \frac{121}{304}e^4}
\frac{\D e}{e},
\end{equation}
and reverse the order of the sum and the integral
\begin{equation}
\left\langle\frac{S^2}{N^2}\right\rangle =
\frac{48}{95} \frac{\eta M_{{\rm tot},z}^3 \rho_{p0}^{2}}{d_L^2}
\int\limits_{e_{\rm LSO}}^{e_{0}}\sum_{n=2}^{n_{\max}(e)}
  \frac{g(n,e)s(e,e_0)}{n^2 S_h(f_n)}
\frac{\D e}{e},\label{sn4-I}
\end{equation}
where $e_{\rm LSO}$ corresponds to the particular $\rho_{p0}$ (see Eq.~\ref{eq:LSO}),
$e_0$ is given by Eq.~(\ref{eq:e}), and
\begin{equation}
s(e,e_0)=\left(\frac{e}{e_0}\right)^{\frac{24}{19}}
\left(\frac{1+\frac{121}{304}e^2}{1+\frac{121}{304}e_0^2}\right)^{\frac{1740}{2299}} \frac{(1+e_0^2)(1-e^2)^{3/2}}{1-\frac{183}{304} e^2 -\frac{121}{304} e^4}.
\end{equation}
Note, that $S_h(f)=\infty$ is assumed outside of $f_{\min}\leq f\leq f_{\max}$.
The upper limit of the sum in this form can be adjusted to the required calculation precision using Eq.~(\ref{nmax}).

For Phase~{\bf II}, we can rewrite Eq.~(\ref{sn4-I}) by changing the sum
over $n$ to a continuous integral over $f_n$.
\begin{equation}
\left\langle\frac{S^2}{N^2}\right\rangle =
\frac{48}{95}\frac{\eta M_{{\rm tot},z}^3 \rho_{p0}^{2}}{d_L^2}
\int\limits_{e_{\rm LSO}}^{e_0}
\int\limits_{f_{\min}}^{f_{\max}}
\frac{g(n,e)s(e,e_0)}{{f_{\rm orb},z} S_h(f)}
\frac{\D f}{f}\frac{\D e}{e}.\label{sn4-II}
\end{equation}
where $n=f/f_{{\rm orb},z}$, $f_{{\rm orb},z}=f_{\rm orb}(r_{p,z},e,e_0)$
given above by Eq.~(\ref{eq:forb}), and $r_{p,z} = M_{{\rm tot},z} \rho_{p0}=(1+z)M_{\rm tot} \rho_{p0}$.

In addition to their numerical advantages, Eqs.~(\ref{sn4-I}) and
(\ref{sn4-II}) can be used to study the time-frequency evolution of the
instantaneous $S/N$ accumulation rate as the orbit evolves. Figure~\ref{f:snr-eccentricity-frequency}
shows the contribution of the first $n\leq 100$ harmonics to the
signal-to-noise ratio for AdLIGO (i.e. before evaluating the sum or the
integral in Eq.~\ref{sn4-I}) for total masses $\mtot=(20,60)\Msun$ and
initial periapse $\rho_{p0}=(10,20,40)$. The figure illustrates the
unique frequency evolution of the signal consistent with the
expectations described above. Initially during Phase~{\bf I}, it is
broadband in frequency, and decouples into discrete harmonics at
smaller eccentricities during Phase~{\bf I}. The contribution of upper
harmonics is nonnegligible even at LSO, especially if the initial
periapse satisfies $\rho_{p0}\lsim 40$. Note that the maximum
frequency of the $S/N$ at large eccentricities $e\gtrsim 0.8$ in
Figure~\ref{f:snr-eccentricity-frequency} is merely a consequence of
not plotting harmonics beyond $n=100$, leading to a large
underestimate of $S/N$ in Phase~{\bf II}. In this case Eq.~(\ref{sn4-II})
becomes more useful than Eq.~(\ref{sn4-I}).

\begin{figure}
\centering
 \mbox{\includegraphics{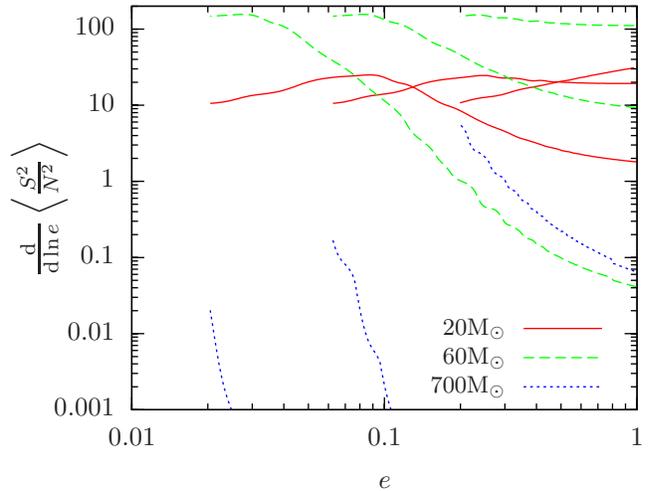}}
 \caption{\label{f:dsnr-eccenticity} The square of the
signal-to-noise ratio per
   logarithmic eccentricity bin as a function of eccentricity for
   AdLIGO at $1\,$Gpc. Three sets of curves are shown for total masses with
   different line types as labelled and initial periapse
   $\rho_{p0}=(10,20,40)$ from right to left corresponding to
   Figure~\ref{f:snr-eccentricity-frequency}. Eccentric encounters
   between massive black holes are detectable up to $700\Msun$ with
   AdLIGO that are totally invisible during a circular inspiral.  }
\end{figure}

The frequency-independent $S/N$ accumulation rate can be obtained as a
function of eccentricity (or time using the $e(t)$ dependence shown in Fig.~\ref{f:t-e}),
if evaluating the sum over $n$ in
Eq.~(\ref{sn4-I}) or the integral over $\D f/f$ in Eq.~(\ref{sn4-II}),
but not the integral over
eccentricity. Figure~\ref{f:dsnr-eccenticity} shows the result for the
same masses and periapse as Fig.~\ref{f:snr-eccentricity-frequency},
including an additional extreme case of $\mtot=700\Msun$. For BH masses
below $\mtot\sim 100$, the $S/N$ contribution of the high and low
eccentricity phases are comparable. The $S/N$ during highly eccentric
encounters dominate for the typical case $\mtot\sim 20\Msun$ for
$\rho_{p0}\lesssim 20$, while small $e$ dominates for larger masses
$\mtot\sim 60\Msun$ for $\rho_{p0}\gtrsim 20$. It is very
interesting that the eccentric encounters between binaries with intermediate
masses up to $\mtot=700\Msun$ are detectable to $\approx 1\,$Gpc for close encounters
$\rho_{p0}\lesssim 10$ with AdLIGO, because these masses are otherwise
totally invisible during a circular inspiral. This is explained by the
broadband nature of the signal during Phase~{\bf I}, leading to
nonnegligible power leaking into the detectable frequency range of the
detector $f\geq 10\Hz$. Such massive BHs were thought to be accessible
only during the violent merger/ringdown phase for AdLIGO \citep{1998PhRvD..57.4535F,2007PhRvD..75l4024B}.

The total $S/N$ can be obtained by evaluating both the sum and the
integrals in Eqs.~(\ref{sn4-I}) and (\ref{sn4-II}). The result can be
converted into a maximum distance of detection assuming a detection
threshold, e.g. $\langle S^2/N^2 \rangle\equiv 5^2$. For Phase~{\bf II},
\begin{equation}
\label{eq:luminositydistance}
d_{L}^{\max} =
\sqrt{\frac{48}{95} \frac{\eta M_{{\rm tot},z}^3 \rho_{p0}^{2}}{\left\langle S^2/N^2 \right\rangle}
\int\limits_{e_{\rm LSO}}^{e_{0}}\sum_{n=2}^{n_{\max}(e)}
  \frac{g(n,e)s(e,e_0)}{n^2 S_h(f_n)}
\frac{\D e}{e}},
\end{equation}
and similarly for Phase~{\bf I}. Note, that the integral depends on two parameters $\mtot$ and $\rho_{p0}$ and is independent of the mass ratio $\eta$.

\begin{figure}
\centering
 \mbox{\includegraphics{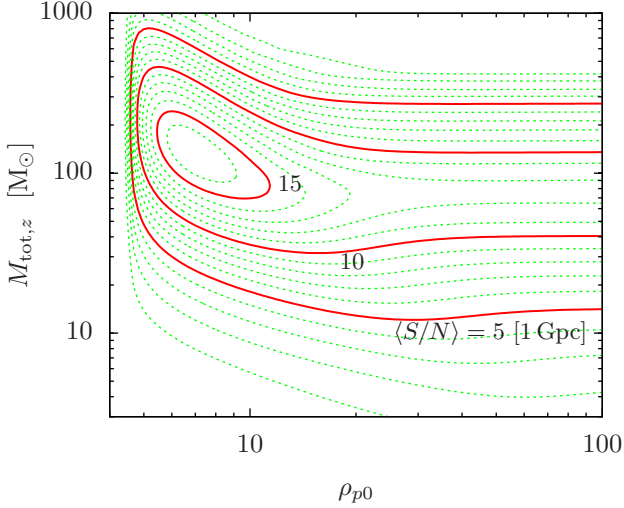}}
 \caption{\label{f:snr} The total signal-to-noise ratio contours as a
   function of the pericenter separation $\rho_{p0}$ of the first passage and total
   binary mass, $M_{{\rm tot},z}$ for AdLIGO for an equal-mass binary at
   $1\,$Gpc. Different contours show $\langle S^2/N^2\rangle^{1/2}$
   (averaged over the binary orientation and sky position) increasing
   in steps of 5 for the solid and 1 for the dotted lines. For other
   mass ratios the SNR is reduced by $(4\eta)^{1/2}$. The limit
   $\rho_{p0}\rightarrow \infty$ corresponds to circular inspirals in
   the AdLIGO band, which result in typically a smaller $S/N$ than
   most likely events with $\rho_{p0}\lesssim 40$. Note that different
   solid contours correspond to a maximum distance of detection
   increasing in steps of Gpc for a detection threshold $S/N=5$.  }
\end{figure}

Figure~(\ref{f:snr}) shows contours of $d_L^{\max}$ for masses $2\Msun
\leq \mtot \leq 1000\Msun$ and initial periapse $4 \leq \rho_{p0}\leq
100$. For the limit of large $\rho_{p0}$, the signal is circularised
when it arrives in the LIGO band, corresponding to the circular
inspiral. The figure shows that the eccentric inspirals can be
detected just to a slightly larger distance than the circular
inspirals for masses $\mtot\leq 20\Msun$. The difference becomes more
pronounced at much larger masses. A 100$\Msun$--100$\Msun$ eccentric
inspiral is detectable up to $3\times$ farther than a standard
$10\Msun$--$10\Msun$ circular inspiral. For even larger masses,
$d_{L}^{\max}$ decreases, but remains nonnegligible up to
$\mtot=700\Msun$. Currently less is known about the existence of such
intermediate mass black holes in galactic cusps, which makes it very
difficult to make any theoretical estimates on the expected rates of
such encounters.  It is possible that AdLIGO will provide the first direct
observational limit on the population of these objects.  We explore
this along with other estimates of the detection rate in \S~\ref{sec:detrate}.

\subsection{Detection Rate Estimates}
\label{sec:detrate}

\begin{figure*}
\centering
 \includegraphics[width=\columnwidth]{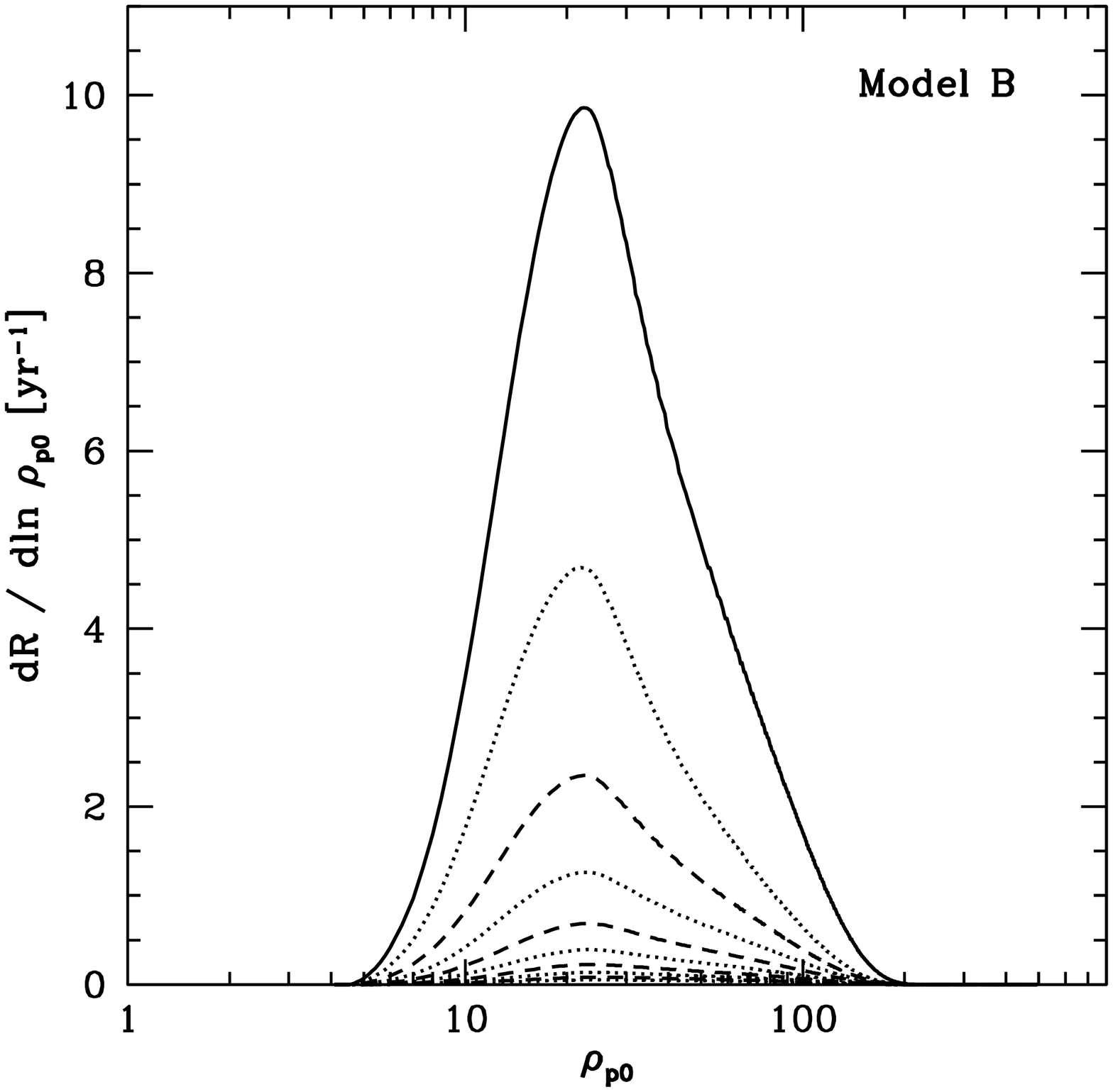}
 \includegraphics[width=\columnwidth]{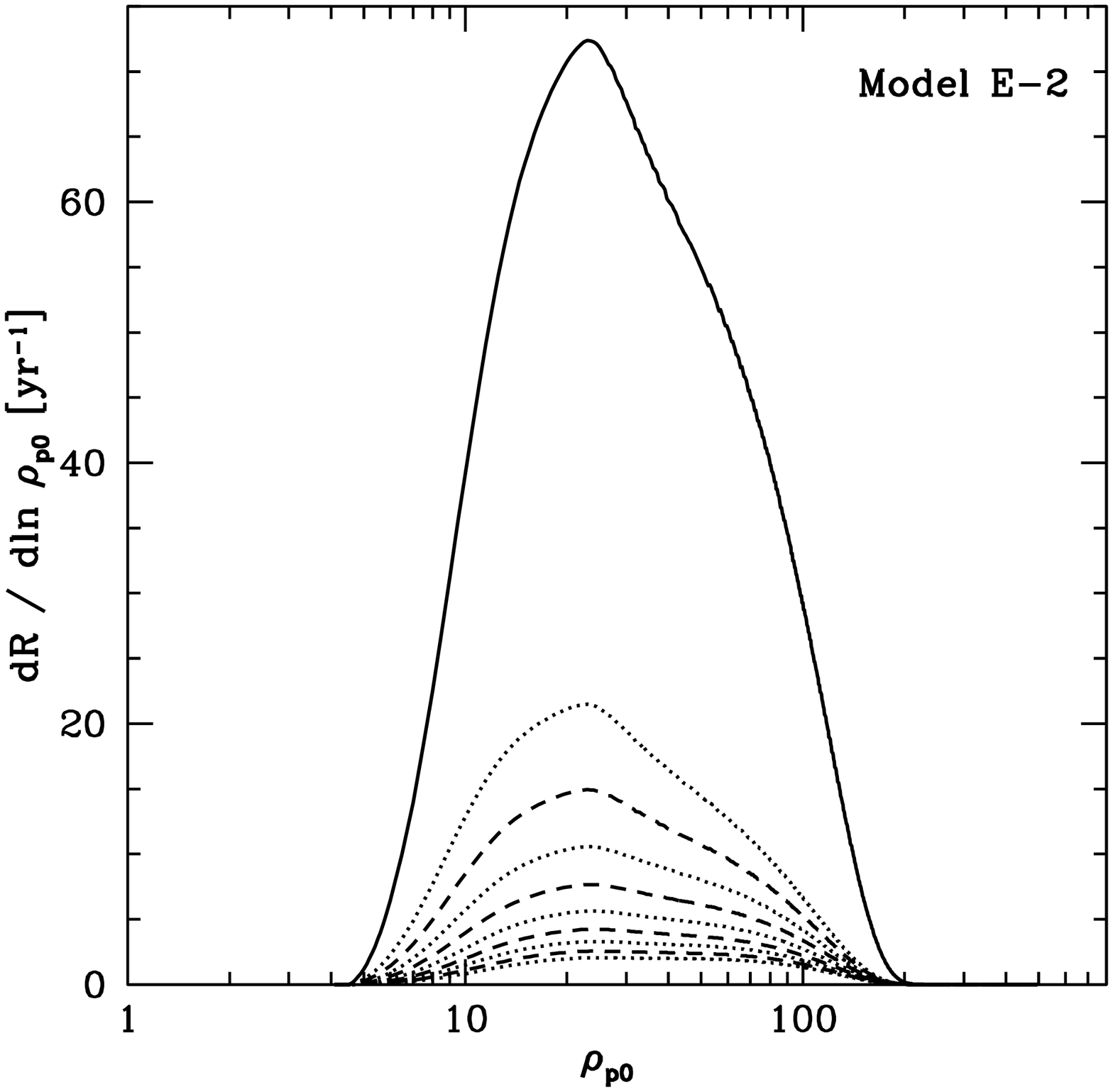}
 \caption{\label{fig:diffrate} The pericenter distribution for AdLIGO
   detections. Plotted is the differential detection rate ($\rmd
   R/\rmd \ln{\rho_{p0}}$) as a function of $\rho_{p0}$ for Models B
   (left) and E-2 (right). In both figures, the solid line is
   differential rate for the all detections, the alternating dotted
   and dashed lines are for each individual mass bin M ($\sum_m \rmd
   R_{mM}/(2 \rmd \ln{\rho_{p0}})$) in order of decreasing mass from
   top to bottom.  Note that function plotted here is different than
   in Fig.~\ref{fig:distribution}}
\end{figure*}

We set $\left\langle S^2/N^2 \right\rangle^{1/2} \geq 5$ as our
detection threshold for a single GW instrument, and determine the volume averaged maximum
luminosity distance for detection for each pair of masses, $d_{\rm
  L}^{\max}(m,M)$ from Eq.~(\ref{eq:luminositydistance}),
assuming a uniform distribution of $r_p$ out to $r_{pmax}$.  Thus, the
total detection rate for Advanced LIGO is estimated to be
\begin{eqnarray}
  \label{eq:totalrate}
   R &=& \int_0^{\rho_{p0}^{\max}} \rmd \rho_{p0} \int\limits_{M_{\min}}^{M_{\max}} \int\limits_{M}^{M_{\max}} \rmd m \rmd M
   \int_0^{z_{\max}} \rmd z \frac{\rmd V_{\rm co}}{\rmd z} \nonumber\\
   \quad&&\times\frac{1}{1+z}\frac{\partial^3 {\cal R}}{\partial \rho_{p0} \partial m \partial M}
\end{eqnarray}
where $\partial^3 {\cal R}/\partial
\rho_{p0}\partial m \partial M$ is the comoving partial binary
formation rate density between masses $(m,M)$ at initial periapse
$\rho_{p0}$, given by Eqs.~(\ref{eq:diffsimpfin}) and (\ref{eq:cosrate}),
averaged over the distribution of $\msmbh$ and the number
density normalisation $n_{*}$ (see Eq.~\ref{eq:smbhnumber} and
\ref{eq:cosrate}) at redshift $z$, $n_{\rm gal}$ is the comoving
number density of galaxies at redshift $z$, $z_{\max}$ is the maximum
detectable distance corresponding to $d_{L\max}(\rho_{p0},m,M)$, and
$\rmd V_{\rm co}/\rmd z$ is the comoving volume density corresponding
to the given cosmology.  In practice, however, we calculate $R$
discreetly and ignore cosmological
effects, which are not relevant for the next generation of GW
instruments,
\begin{eqnarray}
  \label{eq:totalrateapprox}
  R &\approx& \sum_{M,m<M} \int_0^{\rho_{p0}^{\max}} \rmd \rho_{p0}
  \frac{\rmd {\cal R}_{mM}}{\rmd \rho_{p0}}
  \frac{4 \pi}{3} (d_L^{\max})^3,
\end{eqnarray}
where $\rmd {\cal R}_{mM}/\rmd \rho_{p0}$ is the differential rate
density of binary formation between mass bins $m$ and $M$ (see
Eqs.~\ref{eq:diffsimpfin} and \ref{eq:cosrate}).  The resulting total
detection rates are shown in the last column of Table~1. Note that the
dependence of $d_{L\max}$ on the binary parameters $(\rho_{p0},m,M)$
lead to an observational bias. For example, since $d_{L\max}$ is
relatively larger for $\rho_{p0}\sim 10$, the detection rate of these
encounters is enhanced relative to larger and smaller $\rho_{p0}$,
even though the intrinsic rate of these encounters is independent of
$\rho_{p0}$. We have plotted the differential detection rate as a
function of $\rho_{p0}$ in Figure~\ref{fig:diffrate} for all mergers
as well as each mass bin.

\begin{figure*}
\centering
 \includegraphics[width=\columnwidth]{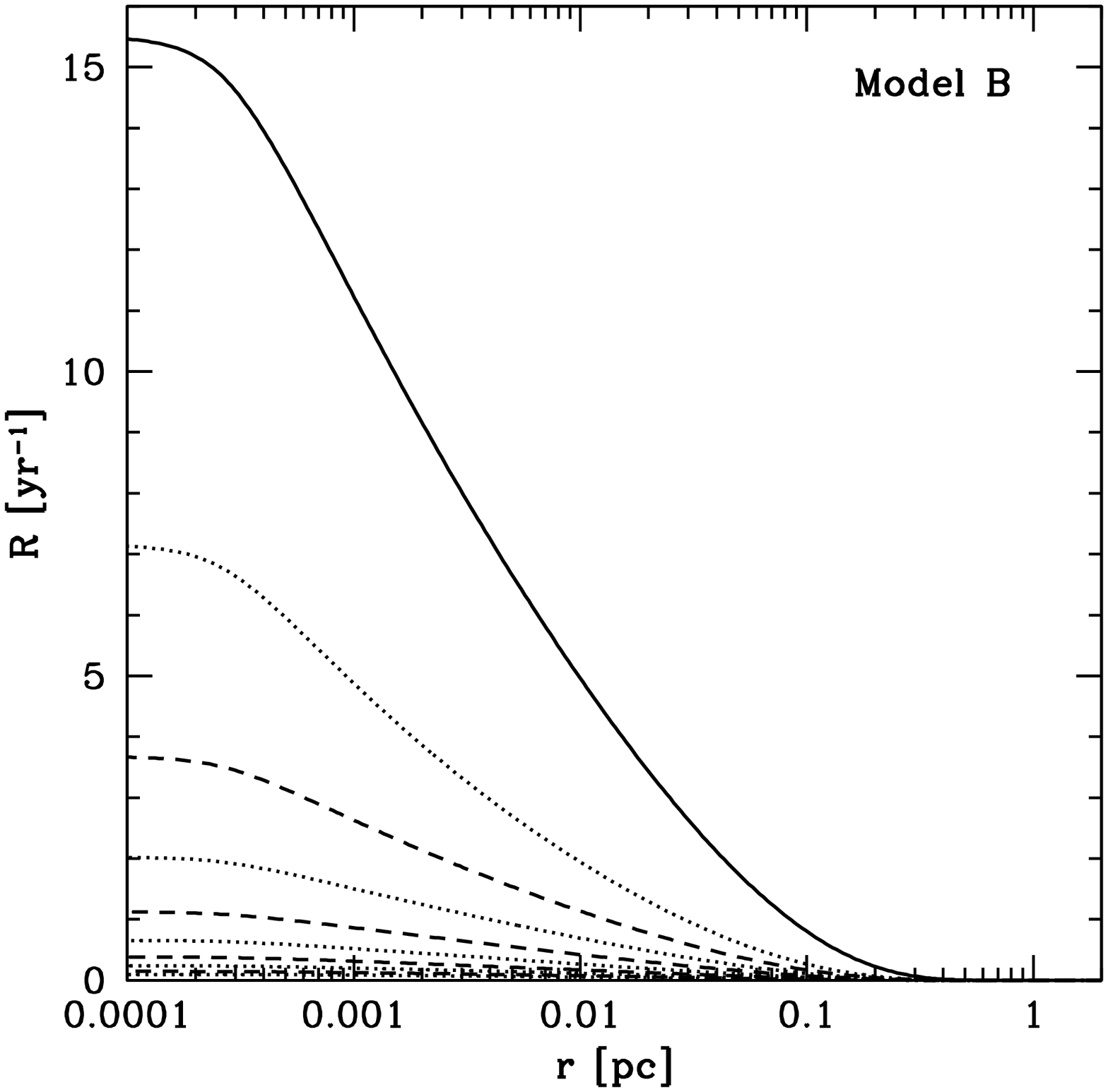}
 \includegraphics[width=\columnwidth]{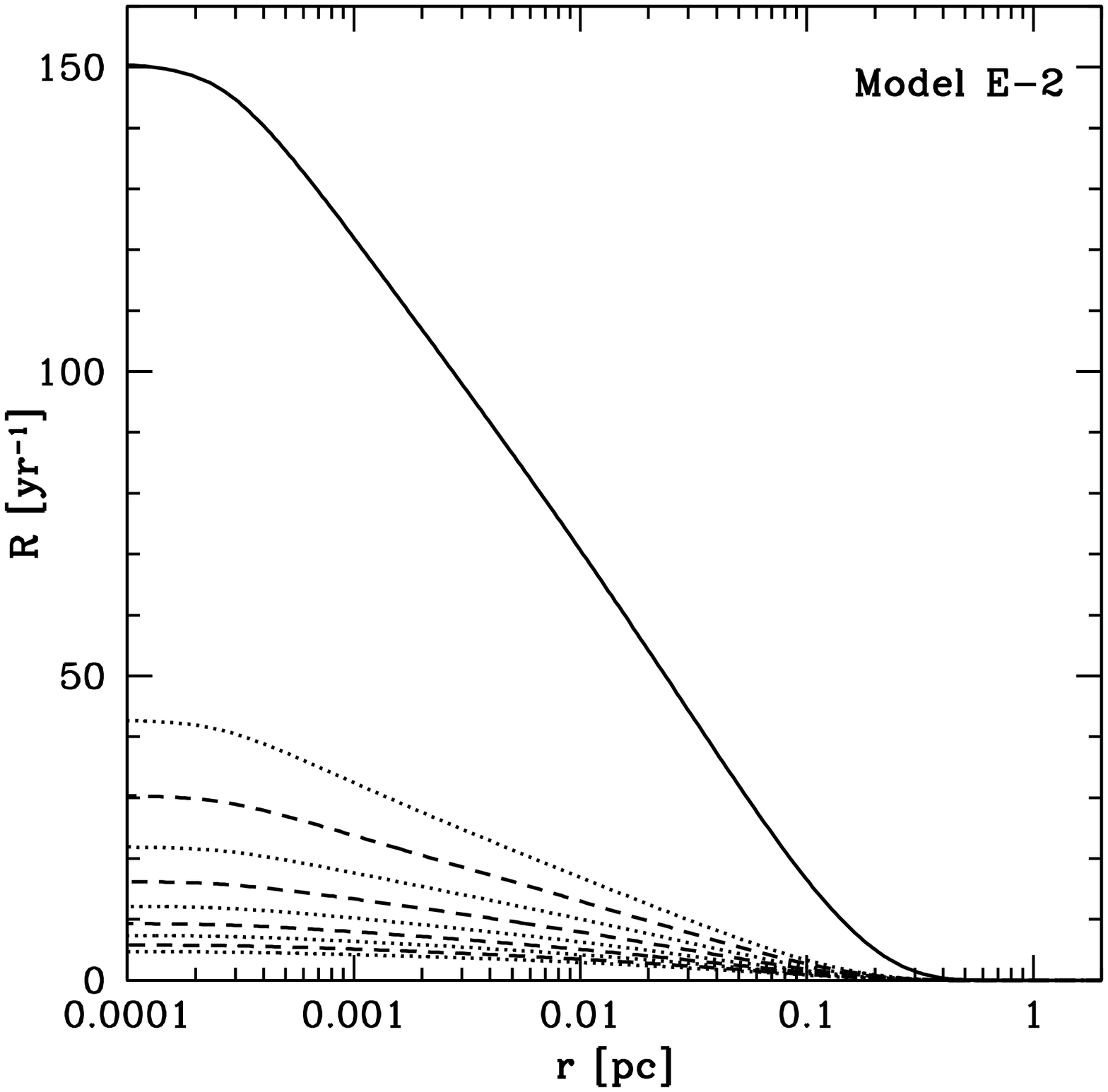}
 \caption{\label{fig:detectionrate} The AdLIGO detection rate for
   Models B and E-2.  The solid line is the total integrated AdLIGO
   detection rate at radii larger than $r$, ($R(>r)$), the alternating dotted and dashed lines
   are the contribution to the detection rate of each individual mass bin ($\sum_m
   R_{mM}(>r)/2$) in order of decreasing mass from top to bottom.}
\end{figure*}

Overall, the most massive BHs in galactic nuclei dominate the
detection rate of mergers for AdLIGO.  In
Figure~\ref{fig:detectionrate}, we have plotted the distribution of
detectable mergers as a function of radius for the entire population
of BHs, as well as for each mass bin.  The clear domination of the
high mass BHs is caused by a combination of three important factors:
{\it (i)} their number density is significantly enhanced by mass
segregation (\S~\ref{sec:massseg}); {\it (ii)} the signal of the event
is much stronger for larger masses (e.g.,
Eqs.~\ref{sn4-I}~\&~\ref{sn4-II}); and {\it (iii)} the cross-section
for binary capture is greater for larger masses (e.g.,
Eq.~\ref{eq:bmax}).

  As discussed in \S~\ref{sec:luminositydistance},
we expect the actual inspiral of a BH with an IMBH may be revealed by
AdLIGO if the event is sufficiently eccentric during plunge.  Although we
cannot properly account for this in our analysis, we can attempt to compare
the rate to what we have done in this work.  For a large mass ratio, $m\gg
M$, the overall cross-section for forming binaries increases roughly as
$b^2 \propto m^{12/7}$.  Compared to $10\,\Msun$ BHs, we expect a
$1000\,\Msun$ BH to have a gravitational wave capture event $\sim 3000$
times as often as a single BH.  However, the number of IMBHs in the region
is very uncertain.  If we take the optimistic number of $\sim 10$ IMBHs in
a single galactic nucleus as a steady state distribution
\citep{2006ApJ...641..319P}, then we expect a comparable total number of
events to $10\Msun$--$10\Msun$ BH-BH inspirals.

Our analysis in \S~\ref{sec:massseg} suffers from inaccuracies for the
most massive and rarest BHs. Equation~\ref{eq:fokkerplanck} was
derived assuming a constant density core (and constant relaxation
timescale) for large $r$ (BW76), which is clearly violated in most
galactic nuclei.  Typically, the total number of BHs in galactic
nuclei decreased with $M_{\max}$. Despite this decrease in number,
we likely cannot extrapolate our calculations to higher mass BHs such
as IMBHs, in order to see what effect they have on flattening the
density profile of stellar mass BHs.

\section{Summary and Discussion}

In this paper, we have analysed two separate problems. First we
determined the multi-mass distribution of BHs in galactic nuclei,
which we then used to analyse the detection rate and merger of
gravitational wave capture binaries.  We
integrated the time-dependent Fokker-Planck equations for a variety BH
mass distributions until they reached a steady state.  We found,
consistent with previous results, that within a relaxation timescale
at the radius of influence of the SMBH, the BHs and stars form
approximately power-law density cusps ($\propto r^{-3/2-p_0 m / M_{\max}}$)
within the radius of influence of the SMBH.  Because the BHs
are more massive than the stars, they have steeper density profiles
($\propto r^{-2}$) and dominate the dynamics and relaxation processes
in the inner $\sim 0.1\,$pc near the SMBH.

In such dense population environments, the probability of close flybys
between two BHs is nonnegligible.
Using our results for the steady-state distribution of BHs,
we calculated the expected rate and distribution of GW
capture binaries in galactic nuclei.  We showed that after forming,
these binaries rapidly inspiral and merge within hours.  Unlike other
sources of merging binaries, the BH binaries in galactic nuclei form
with a characteristic GW frequency {\em inside} the AdLIGO
frequency band, and the detectable GW signal duration is much longer
compared to circular inspirals.  In addition, for sufficiently small impact parameters,
the waveforms for such eccentric inspirals are broadband, and can be
detected for much larger masses, up to $\mtot \gtrsim 700\,\Msun$.
This exceeds the BH masses previously considered
detectable \citep{2007PhRvL..99t1102B,2008ApJ...681.1431M}, and as
such it opens a new avenue to probe for the existence of a population of
intermediate mass black holes in galactic nuclei.

Additionally in \S~\ref{sec:clusters}, we also estimated the rate of
GW captures in young, massive star clusters where BHs decouple
from stars to form a subcluster in the centre.
We found that the rate of GW captures in a single cluster may be
intrinsically larger than in a galactic nucleus. Typically, most
binaries circularise before becoming detectable, however a significant
fraction {\new ($\sim 10\%$) }may still merge with residual eccentricity.  Given the vastly
different eccentricity distribution of mergers in galactic nuclei and
massive star clusters, the dominate source of eccentric inspirals can
be determined with only a few detections.  The total rate of such
events from massive clusters depends on the relatively unconstrained
number of young clusters within the detection limits of AdLIGO.

Finally, we analysed the properties of the
GW signal for the eccentric inspiral of comparable mass BHs, starting
from the initial highly eccentric phase up to the final eccentric
inspiral.  We found that the maximum luminosity distance that such
events may be detected using a single AdLIGO-type interferometer is
$1.2\,$Gpc for component masses $m=M\sim 10\,\Msun$, and up to
$3.3\,$Gpc for $m=M\sim 50\,\Msun$.  We then used this to calculate
the total detection rate of signals, by using a model merger rate
convolved with a realistic population of SMBHs.  We found that the
most massive BHs dominate the detection rate of future ground based
gravitational wave detectors with $\sim (10 - 10^3) \xi_{30}$ events
expected per year, where $\xi_{30}$ is a measure of the expectation
value for the square of number densities near the SMBH (see
Eq.~\ref{eq:xi}).
The overall detection rate is sensitive to
the number fraction of BHs as well as the maximum mass of BHs, and so
future observations will be able to constrain both the average star
formation properties and upper mass of BHs in galactic nuclei.

Overall, the expected AdLIGO detection rate for GW capture binaries in
galactic nuclei is comparable to estimates for other sources of
gravitational waves. BH-BH binaries that form dynamically in massive
star clusters may be detected by AdLIGO $\sim 10-10^4\,$ times per
year, depending on the total fraction of star formation that occurs in
massive, long lived clusters \citep{2000ApJ...528L..17P,
  2004ApJ...616..221G, 2006ApJ...637..937O, 2006ApJ...640..156G,
  2007PhRvD..76f1504O, 2008arXiv0804.2783M}. Estimates for the
population of merging binaries that formed in the early evolution of
present day globular clusters are $\sim 10-100\,$yr$^{-1}$ alone.
Although globular clusters contain only a small fraction of the mass
of stars in the universe, the merger rate of BHs is sufficiently
enhanced by dynamical interactions that the rate is comparable to those
expected from standard binary evolution for all other stars
(\citealt{2002ApJ...572..407B, 2004ApJ...611.1068B,
  2008ApJ...676.1162S}; however see \citealt{2007ApJ...662..504B}).
Merging neutron stars are also a promising source of gravitational
waves, whose progenitors have been directly observed.  Using
constraints from the modest population neutron star-neutron star
binaries known in our galaxy,
\citet{2004ApJ...601L.179K,2004ApJ...614L.137K} estimate that $\sim 20
- 600$ events per year may be detected by AdLIGO.  We emphasise that
the rates of GW captures calculated in this paper are {\em
  independent} of other sources of GWs, and will occur in addition to
other sources.  Given the estimates of sources from a variety of
environments, gravitational wave detectors promise to open a new and
interesting window into the physics, dynamics, and evolution of
compact binaries.

So far, we have not looked at the interaction between the BHs and
other compact objects such as neutron stars (NSs). The NSs are not
expected to segregate significantly in galactic nuclei, however they
are intrinsically more common than BHs.  To test their importance we
have performed an additional run of Model E-2 to look at the expected
detection rate of NS-BH mergers. We found that the total detection
rate of such events is about $\sim 1\%$ of the total rate ($\sim
1\,$yr$^{-1}$), and merits future study. Gravitational waves from
BH-NS inspirals can provide interesting constraints on the equation of
state for NSs \citep{2002PhRvL..89w1102F}, given model sources from
numerical simulations \citep{2006PhRvD..73b4012F,2008ApJ...680.1326R}.
To date, such analyses have focused on the circular inspiral of BHs
with NSs, however recent population synthesis studies suggest the
number of BH-NS binaries that form may be as rare as those expected
here, $\sim 1\,$yr$^{-1}$ \citep{2007ApJ...662..504B}.  Given that GW
capture binaries are predominately eccentric throughout inspiral, we
also expect the newly formed BH-NS binaries to be eccentric. However,
with BH-NS binaries, the tidal effects of the BH on the NS can provide
an additional source of energy dissipation in the binary, which can
enrich the GW signal, and cause it to deviate from the point-mass
approximations calculated here.  Future numerical simulations with
eccentric encounters must be done in order to test the importance of
eccentricity on the GW signal for BH-NS binaries.

Another area we neglected is the influence of
binarity on the merger rate of BHs in galactic nuclei. However, we can
estimate their overall importance by looking at what fraction of
binaries are hard enough to survive in galactic nuclei.  Given the
distribution of BH binaries in \citet{2004ApJ...611.1068B}, only a
small fraction, $\lesssim 0.1\%$ of the BH-BH binaries are
sufficiently hard (with an orbital period $\lesssim 10\,$days) not to
be disrupted by repeated encounters with stars or BHs. Interestingly,
this fraction is comparable to the fraction of BHs that merge due GW
capture. The properties of the
merger of such binaries have not been explored yet, however, given the
evidence that such encounters produce few eccentric events in massive
star clusters, we expect that they will not have many eccentric
mergers in galactic nuclei. Unfortunately, the methods we have used
here are not suitable for looking at higher order $N$-body
interactions (such as exchanges, or binary-binary scattering);
simulations similar to FAK06 are more adequate for this type of study.
The rates we calculated might be enhanced if high-dispersion
environments of protogalactic cores without a SMBH also exist,
although such systems are old and not long lived
\citep{1987ApJ...321..199Q,1989ApJ...343..725Q,1990ApJ...356..483Q,1993ApJ...418..147L}.

{\new

  In order to estimate the detection rate of gravitational wave
  capture binaries in galactic nuclei we had to make assumptions that
  introduced uncertainty in our calculations. For the sakeof clarity, we wish
  to summarise these uncertainties and the order of magnitude effect
  they may have on the actual detection rate for AdLIGO.  The largest
  uncertainty in the rate calculation is the distribution of BHs in
  galactic nuclei, especially the number fraction of BHs ($\sim 10^2$)
  and their mass distribution ($\sim 10$) in the centres of galaxies.
  Observations of our own galactic centre suggest that the star
  formation may be top-heavy, and place the rate on the higher end of
  that reported here.  One effect we neglect in our calculations is
  resonant relaxation, which may reduce the rate $(\sim 10)$ by
  depleting the central cusp of BHs. Our final rate, however, is in
  practice only logarithmically sensitive to the inner edge of the
  cusp of BHs, and the densest cusps that dominate our rate will be
  least affected by resonant relaxation. The final rate is also nearly
  logarithmically dependent on the population of SMBHs in the local
  universe (by a factor of at most $\sim 10$), 
but is much more sensitive to the distribution
  of densities at the radius of influence of the BHs (by a factor of $\sim
  10$--$10^2$). Finally, there is some uncertainty in the detection rate
  given our calculation of the $S/N$ of the mergers ($\sim 10$) since
  we do not consider corrections for zoom-whirl orbits and
  conservatively ignore the contribution of the merger and ringdown
  waveforms of the binary to the signal.

}

There are many aspects to the merger of GW capture binaries that
require future work, which will be aided by advanced numerical
simulations of the evolution of these binaries.  In our calculations
we used only the leading order (Newtonian) formula to calculate the evolution
of the binaries and the GW waveforms, and
do not account for the GW signal during the plunge and merger of
the binary, nor do we consider the contribution of zoom-whirl orbits.
The recent breakthrough of numerical simulations
finally allows the full GW signal to be calculated directly. Eccentric
mergers promise to have a richer signal than that for circular
binaries. Future studies should incorporate the contribution of
the final coalescence to the GW signal, which would increase the
signal strength generated during the inspiral for the same capture event,
and has a potential to further increase the maximum range of detection.

We also found that a fraction ($\sim 10-20\%$) of mergers will occur
very close to, or within the estimated last stable orbit of BH
binaries.  These capture events may be zoom-whirl orbits, which
dramatically increases the GW signal compared to a regular inspiral
\citep{2007CQGra..24...83P}.  Such encounters are inherently in the
strong GR regime, and again require a full numerical treatment of
General Relativity to be studied. A precise assessment of these
sources will further increase our estimates of detection rates.  Our
estimates for the number of GW capture binaries in galactic nuclei (and
perhaps in massive star clusters) suggest that such events may be
common enough to be directly detected, and may provide the richest
gravitational wave sources in the universe for ground based detectors.

\section*{Acknowledgements}
We thank Krzysztof Belczynski, Sam Finn, Scott Hughes, Pablo Laguna,
Ilya Mandel, Szabolcs M\'{a}rka, and Coleman Miller for useful
discussions and Andr\'{a}s P\'{a}l for helping with Figure 9.
This work is supported by in part by NASA grant NNX08AL43G, by FQXi, and by
Harvard University funds. BK acknowledges support from OTKA Grant 68228 and
Pol\'anyi Program of the Hungarian National Office for Research and Technology
(NKTH).

\bibliography{p}

\end{document}